\documentclass[11pt,letterpaper]{article}
\pdfoutput=1
\usepackage{jheppub}
\usepackage{bbm}
\usepackage{mathrsfs}
\usepackage{slashed}
\usepackage{caption}
\usepackage{epstopdf}
\usepackage[normalem]{ulem}
\usepackage[bottom]{footmisc}
\usepackage{subcaption}
\usepackage{bbold}
\usepackage{titlesec}
\usepackage{threeparttable}
\usepackage{booktabs}
\usepackage{changepage}
\usepackage[utf8]{inputenc}
\usepackage{dsfont}

\usepackage{grffile}
 
\usepackage{graphicx}  
\usepackage{dcolumn}   
\usepackage{bm}        
\usepackage{amssymb}   
\usepackage{setspace}
\usepackage{amsmath, amssymb, setspace}
\usepackage{array}
\usepackage{booktabs}
\usepackage{caption}
\usepackage{indentfirst}
\usepackage{float}
\usepackage{lmodern}
\usepackage{multirow}
\usepackage{soul}
\usepackage[normalem]{ulem}

\usepackage{braket}
\usepackage{comment}

\usepackage[draft]{pgf}

\usepackage{adjustbox} 
\def \g {\gamma}

\newlength{\myimageoversize}
\newsavebox{\myimage}

\usepackage[most]{tcolorbox}

\usepackage{empheq}

\tcbset{colback=white!10!white, colframe=black!50!black, 
        highlight math style= {enhanced, 
            colframe=black,colback=white!10!white,boxsep=0pt}
        } 

\titleformat{\subsection}
 {\normalfont\fontsize{12}{17}\itshape}{\thesubsubsection}{1em}{}

\title{\huge{Multimessenger Constraints on Supermassive Dark Stars and Their Black Hole Remnants}}

\author[a,b,c]{Marco Manno,}
\author[c]{Thomas Schwemberger,} 
\author[c,d,e,f]{and Volodymyr Takhistov} 

\affiliation[a]{Dipartimento di Matematica e Fisica ``Ennio De Giorgi'', Università del Salento, Italy} \affiliation[b]{INFN- Istituto Nazionale di Fisica Nucleare, sezione di Lecce, Italy}
\affiliation[c]{International Center for Quantum-field Measurement Systems for Studies of the Universe and Particles (QUP,WPI), High Energy Accelerator Research Organization (KEK), Oho 1-1, Tsukuba, Ibaraki 305-0801, Japan}
\affiliation[d]{Theory Center, Institute of Particle and Nuclear Studies (IPNS),
High Energy Accelerator Research Organization (KEK), Tsukuba 305-0801, Japan} 
\affiliation[e]{Graduate University for Advanced Studies (SOKENDAI), \\
1-1 Oho, Tsukuba, Ibaraki 305-0801, Japan}
\affiliation[f]{Kavli Institute for the Physics and Mathematics of the Universe (WPI), Chiba 277-8583, Japan}

\emailAdd{marco.manno@unisalento.it} 
\emailAdd{tschwem2@post.kek.jp}
\emailAdd{vtakhist@post.kek.jp}

\abstract{
Dark matter (DM) annihilation can power the first generation of stars as long lived dark stars (DSs) that grow to supermassive scales $M_{\rm DS}\gtrsim 10^{5} M_{\odot}$  and eventually collapse into heavy black holes that could seed the supermassive black holes observed at high redshifts. We compute the diffuse electromagnetic emission from a cosmological population of such supermassive DSs and their black hole remnants, tracking the entire DS history and including thermal surface radiation, DM annihilation in adiabatically contracted halos as well as late-time emission from DM overdensity spikes around the resulting black holes. After accounting for photon attenuation, we find that DS related contributions can exceed the Fermi-LAT extragalactic $\gamma$-ray background for thermal relic annihilation cross-sections and DM masses below $\sim 1$~TeV. Our results constitute the first population integrated diffuse multimessenger constraints on supermassive DSs as progenitors of early black holes  and demonstrate that diffuse photon and neutrino backgrounds offer a powerful and complementary avenue for probing the role of DM in the formation of the earliest massive structures.
}

\begin{document}
\preprint{KEK-QUP-2025-0024, KEK-TH-2779}
 \maketitle

\section{Introduction}

The first stars are considered to have played a central role in early cosmic evolution, reionization and element enrichment. These low-metallicity stars, known as population~III (Pop~III), are thought to originate from collapse of pristine primordial gas in dense dark matter (DM)  $\sim 10^{5}$-$10^{7} M_\odot$ halos at redshifts of around $z\sim20$~\cite{Abel:2001pr,Bromm:2001bi,Yoshida:2006bz}.
Despite their significant expected influence on the early Universe direct observational evidence for Pop III stars remains elusive, although several high redshift candidates have been suggested~\cite{vanzella:2020,welch:2022,wang:2024}.

Another key role in cosmic evolution is though to have been played by supermassive black holes (SMBHs), with $\sim 10^6- 10^9 M_{\odot}$ SMBHs populating centers of galaxies, powering quasars and active galactic nuclei. Detection of SMBHs at high $z \gtrsim 10$ redshifts with recent James Webb Space Telescope (JWST) observations~\cite{Matthee:2023utn,Yue2024,Bogdan:2023ilu,Ding:2023,stone:2024} challenge conventional formation scenarios based on Eddington-limited accretion or hierarchical mergers of stellar remnants~\cite{Venemans:2013npa, Banados:2017unc}.
 While Pop~III stars can account for lighter  SMBH progenitors, heavier $\sim10^{4}$-$10^{6}~M_\odot$ SMBH seeds instead suggest alternative formation channels such as direct collapse~\cite{Volonteri:2010wz,Inayoshi:2019fun}.

The elevated DM densities in the early Universe can carry significant implications for stellar formation and could lead to a novel class of earliest  stars 
 powered primarily by DM heating rather than nuclear fusion and that could seed SMBHs, known as dark stars (DSs)~\cite{Spolyar:2007qv,Freese:2008wh}. 
Such heating can be a general consequence of DM annihilation in variety of theories, such as scenarios with weakly interacting massive particles (WIMPs)~\cite{Spolyar:2007qv}. 
During
protostellar collapse, baryonic infall adiabatically contracts the surrounding DM. This increases the central DM density and enables efficient DM  annihilation.
The resulting energy injection from DM annihilation halts collapse at low temperatures $T \lesssim 10^4$~K with suppressing stellar winds, allowing prolonged accretion phases and enabling DSs to grow to supermassive  $\gtrsim10^5~M_\odot$ scales at high redshifts~\cite{Freese:2010re}. DSs can potentially be directly observed~\cite{Freese:2010re} and several possible candidates have been recently put forth using JWST data~\cite{Ilie:2023zfv,Ilie:2025zzj}. Once the supply of DM heating is exhausted, these objects are expected to undergo gravitational instability that results in their dynamical collapse to massive BHs that can act as progenitors of SMBHs. Detecting DSs and their signatures would thus carry significant implications for cosmology, the origin of SMBHs and also the nature of DM.

In this work, we investigate the diffuse multimessenger emission produced by a cosmological population of supermassive DSs and their post-collapse BH remnants, focusing on the electromagnetic contributions. This analysis complements recent work by some of us showing that diffuse neutrino emission from DM powered massive DSs can already be probed with existing neutrino telescopes such as Super-Kamiokande and IceCube, yielding strong limits on such populations\footnote{
Neutrino emission from collapsing supermassive stars can also produce detectable signals in DM experiments~\cite{Munoz:2021sad}.}.
Earlier diffuse emission studies focused primarily on smaller Pop III-scale DSs with mass $\sim10^{2}$-$10^{3} M_\odot$  and considered only DM annihilation luminosity during the DS phase~\cite{Schleicher:2008gk, Schleicher:2008aa, Maurer:2012fv, Yuan:2011yb}.
Supermassive DSs, thermal stellar emission, post-collapse BH-spike emission  or a self consistent link to the observed SMBH population were not considered.
Our work addresses all these elements within a unified framework and establishes the first diffuse, population-integrated multimessenger constraints on supermassive DSs as progenitors of the earliest massive black holes.

This work is organized as follows. In Sec.~\ref{sec:dsevol}
we discuss DS evolution and emission, including emission associated with the pre-collapse DS phase and post-collapse BH phase. Sec.~\ref{sec:ds_population} we discuss the DS population related to DM halos and relate it to abundance of SMBHs. Sec.~\ref{sec:flux} includes computation and analysis of combined diffuse emission flux from population of supermassive DS. In Sec.~\ref{sec:constraints} we compare the DS emission flux to experimental observations and set new constraints on the associated DM parameters. We conclude in Sec.~\ref{sec:conclusions}.

\section{Dark Star Evolution and Emission}
\label{sec:dsevol}

DS evolve in a qualitatively different way from conventional Pop~III stars. Their lower interior densities and temperatures suppress efficient nuclear burning and permit sustained accretion. This enables them to grow to supermassive scales. Once the supply of DM fueling is no longer efficient, these objects are expected to directly collapse to BHs.

\begin{figure}[t]
    \centering
    \includegraphics[width=0.8\linewidth]{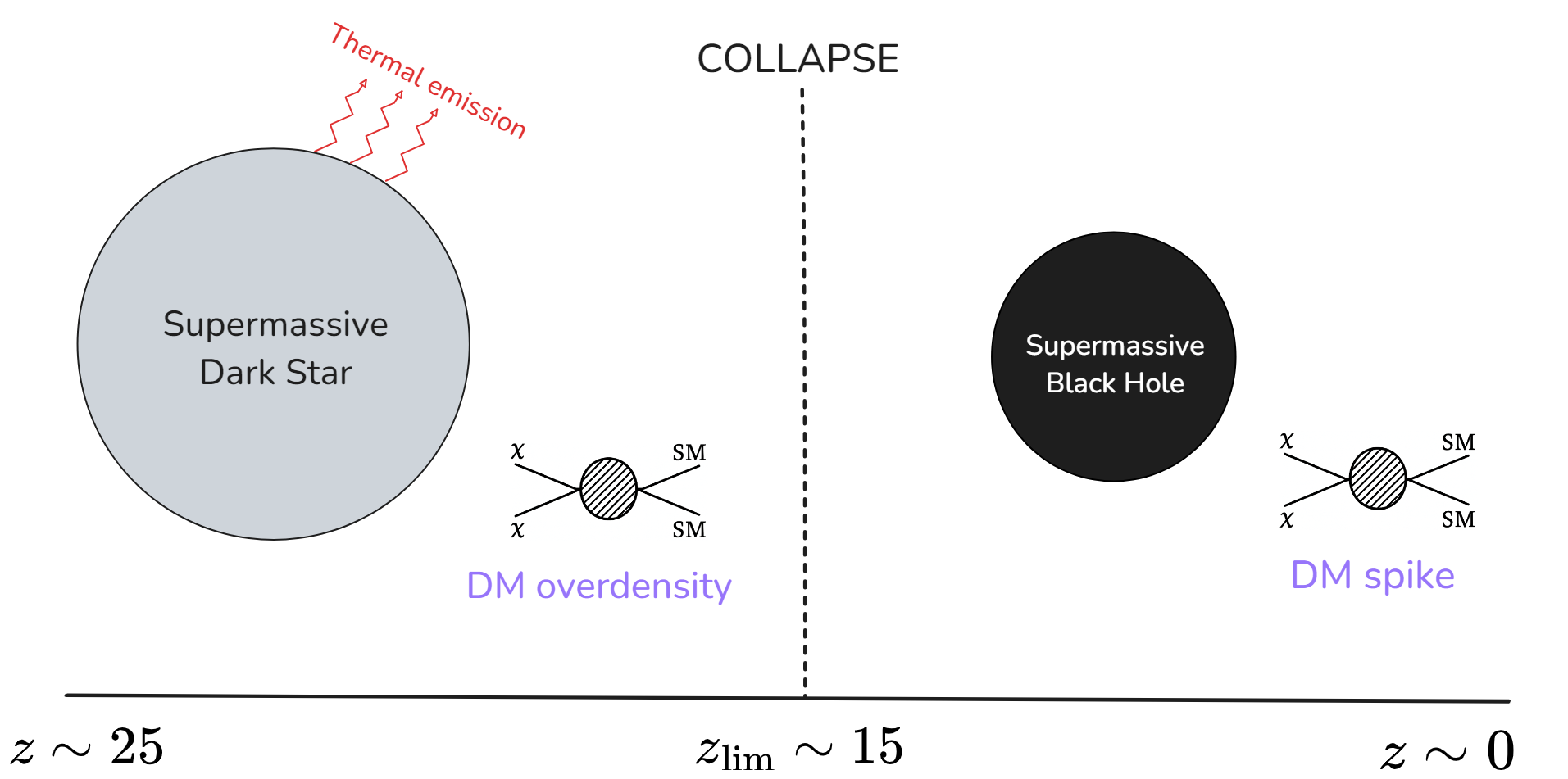}
    \caption{Illustration of the considered electromagnetic emission components. A supermassive DS forming at high redshift $z \sim 25$ emits thermal radiation from its surface and is embedded in an adiabatically contracted DM overdensity, where DM annihilations outside the DS generate an additional non-thermal component. At a characteristic redshift $z_{\rm lim} \sim 15$ (vertical dashed line), the DS exhausts its DM fuel and gravitationally collapses into a SMBH. The resulting SMBH is surrounded by a DM density spike from the progenitor halo, with DM annihilations in the spike producing a non-thermal emission that persists from the time of collapse to low redshifts $z\to0$.}

    \label{fig:DS_evolution}
\end{figure}

We compute the emission associated with DS considering several distinct contributions. First, we treat DM annihilations within the baryonic interior of the star, where the resulting energy is efficiently trapped and ultimately emerges as thermal radiation from the stellar surface. Next, we account for DM annihilations in the adiabatically contracted halo external to the stellar radius, where the produced photons can escape directly. Finally, after the DS collapse, we consider DM annihilations in the contracted DM halo resulting in an overdensity ``spike'' surrounding the BH. The sequence of these stages at different redshifts $z$ is illustrated schematically in Fig.~\ref{fig:DS_evolution}.

The expected thermal DS luminosities have motivated recent individual event searches in JWST data~\cite{Ilie:2023zfv}. Here, we focus on DS populations that persisted at earlier redshifts. As the exact details of DS collapse are complex and an area of active investigation we model that DSs rapidly collapse to massive BHs at redshift $z_{\rm lim} = 15$ as indicated by the vertical dashed line in Fig.~\ref{fig:DS_evolution}. A more gradual population transition in the number of surviving objects, as discussed in Ref.~\cite{Schwemberger:2024rvu} and that can appear from distinct astrophysical environments and different initial populations, could result in a small remaining DS population at $z\lesssim15$  potentially accommodating some of the suggested DS candidates~\cite{Ilie:2023zfv}.

Throughout this work, stellar structure and emission are computed in CGS units, while DM related quantities such as masses, cross-sections and densities, including DM halo and spike densities, are treated in natural units (e.g. GeV, cm$^{3}$~s$^{-1}$, GeV~cm$^{-3}$). For luminosity calculations we convert them as appropriate.

\subsection{Stellar evolution and luminosity}\label{sec:polytrope}

We model DSs as polytropes in hydrostatic and thermal equilibrium to analyze their evolution. This analytic treatment agrees within a factor of a few with 1D stellar evolution calculation results obtained with MESA over the radii and luminosities of interest~\citep{Rindler-Daller:2014uja}. Powered by DM heating rather than nuclear fusion, DSs remain cooler and more diffuse than main-sequence stars. In the gas pressure-dominated phase we adopt a polytropic index $n=3/2$~\citep{Freese:2008wh}. At later stages that are dominated by radiation $n=3$ model has also been previously considered~\citep{Spolyar:2009nt}. We have verified that using $n=3$ decreases the stellar radius and raises the temperature relative to $n=3/2$ changing luminosity $L$ by only $\mathcal{O}(1)$. Since our results are not strongly sensitive to the precise temperature or radius, we adopt $n=3/2$ throughout.

The equation of state relates pressure $P$ and density $\rho$ as
\begin{equation}\label{eq:poly}
    P = K \rho^{1+1/n} ,
\end{equation}
with the normalization $K$ set by the outer boundary conditions. Hydrostatic balance for enclosed mass
\begin{equation}
    M_r = 4\pi\!\int_0^r\!dr' r'^2\rho(r')
\end{equation}
implies
\begin{equation}\label{eq:hydro_eq}
    \frac{dP}{dr} =- \rho \frac{G M_r}{r^2} ,
\end{equation}
where $G$ is the gravitational constant.
Eq.~\eqref{eq:poly}-\eqref{eq:hydro_eq} reduce to the Lane-Emden form
\begin{equation}\label{eq:lane_em}
    \frac{1}{x^2}\frac{d}{dx}\!\left(x^2\frac{d\theta}{dx}\right) =- \theta^{n} ,
\end{equation}
for $x=r/\alpha$, $\rho=\rho_{\rm CD} \theta^n$ with $\rho_{\rm CD}$ being the central density, $\alpha=\big[(n+1)K/4\pi G\big]^{1/2}\rho_{\rm CD}^{(1-n)/2n}$, with central boundary conditions $\theta(0)=1$, $\theta'(0)=0$, and surface at $x_1$ where $\theta(x_1)=0$ ($x_1\simeq3.65$ for $n=3/2$). For a DS with mass 
 $M_{\rm DS}$ and radius $R_{\rm DS}$ the normalization constant is
\begin{equation}\label{eq:K}
    K = \frac{1}{n}
        \left(\frac{R_{\rm DS}}{x_1}\right)^{-1+3/n}
        \left(\frac{G M_{\rm DS}}{-x_1^2\theta'(x_1)}\right)^{1-1/n}
        (4\pi G)^{1/n}.
\end{equation}
For the average density $\bar\rho = M_{\rm DS}/(4\pi R_{\rm DS}^3/3)$ the density
contrast $D_n = \rho_{\rm CD}/\bar\rho$ fixes $\rho_{\rm CD}$ (for $n=3/2$, $D_{3/2}=5.99$).

Thermal support includes both gas and radiation pressure contributions
\begin{equation}\label{eq:EoS}
    P(r)=\frac{\rho k_B T(r)}{\bar m}+\frac{4}{3}\frac{\sigma_{\rm B}}{c}T(r)^4 ~.
\end{equation}
 Here the Boltzmann constant is 
$k_B = 1.38\times10^{-16}~\mathrm{erg~K^{-1}}$, the Stefan-Boltzmann 
constant is $\sigma_{\rm B} = 5.67\times10^{-5}~\mathrm{erg~cm^{-2}~s^{-1}~K^{-4}}$, 
and $c$ is the speed of light. 
The mean particle mass is $\bar m = 0.588m_u$, where the atomic mass unit is 
$m_u = 1.66\times10^{-24}~\mathrm{g}$, corresponding to primordial composition 
$X=0.76$ and $Y=0.24$. The resulting photosphere radius $R_s$ can then be obtained by solving
\begin{equation}
    P \kappa \big(\rho(R_s),T(R_s)\big)=\frac{2}{3} g(R_s) ,
\end{equation}
where $g(r)=GM_r/r^2$ is the gravitational acceleration and the opacity $\kappa$ is taken from the OPAL opacity data~\citep{1996ApJ...464..943I}. This yields the effective temperature $T=T(R_s)$ and the surface luminosity
\begin{equation}\label{eq:surface_lum}
    L_{\rm surf}=4\pi\sigma_{\rm B}R_s^2 T(R_s)^4 .
\end{equation}

The total energy budget includes nuclear fusion, gravitational, and DM heating contributions 
\begin{equation}\label{eq:Lums}
    L_{\rm tot}=L_{\rm fus}+L_{\rm grav}+L_{\rm DM}\simeq L_{\rm DM} ,
\end{equation}
where the last term dominates during the DS phase, while neutrinos escape freely~\citep{Spolyar:2007qv}. The heating from DM annihilation is
\begin{equation}\label{eq:L_dm}
    L_{\rm DM}\simeq (1-f_\nu)\int_0^{R_s}dr 4\pi r^2 
    \frac{\langle\sigma v\rangle}{m_\chi} \rho_\chi(r)^2 ,
\end{equation}
where $m_\chi$ is the DM particle mass and $\langle \sigma v \rangle$ is the thermally averaged annihilation cross section. The integrand represents the local energy injection rate for two-particle DM annihilation, proportional to $\rho_\chi^2(r)$. For Majorana DM the rate receives an additional factor of two because particle and antiparticle populations coincide.

The factor $f_\nu$ denotes the fraction of DM annihilation energy carried away by neutrinos, which free stream out of the star without contributing to heating. Electromagnetic and hadronic annihilation products thermalize efficiently, and thus only the fraction $(1-f_\nu)$ contributes to the stellar luminosity. A detailed discussion and calculation of $f_\nu$ is provided in App.~\ref{app:neutrinos}.

We model the contracted DM profile anchored to the baryonic core following Ref.~\citep{Spolyar:2007qv} as
\begin{equation}\label{eq:dm_dens}
    \rho_\chi(r)\simeq \rho_{\chi,0}\left(\frac{r}{r_0}\right)^{-1.9},
    \qquad
    \rho_{\chi,0}\simeq 1.7\times10^{11}\ \mathrm{GeV ~cm^{-3}}
    \left(\frac{n_b}{10^{13}\ \mathrm{cm^{-3}}}\right)^{0.81},
\end{equation}
where $\rho_{\chi,0}$ is the DM density at the outer edge of baryonic core after adiabatic contraction, $r_0$ is the outer radius of the baryonic core and $n_b$ is the baryon number density. We consider $n_b=10^{13}~   \mathrm{cm^{-3}}$ at DS formation~\citep{Spolyar:2007qv}. We then iteratively solve the polytrope equations and as the DS evolves the $n_b$ changes accordingly. Outside the core  we adopt the representative DM profile $\rho_\chi\propto r^{-1.9}$.

In thermal equilibrium, the DM heating rate of Eq.~\eqref{eq:L_dm} balances surface emission of Eq.~\eqref{eq:surface_lum} giving
\begin{equation}\label{eq:theq}
    L_{\rm DM} \simeq L_{\rm surf} .
\end{equation}
Procedurally, for a given $M_{\rm DS}$ we 
(i) solve the polytrope for a trial radius $R_{\rm DS}$ finding corresponding temperature profile, (ii) determine photosphere radius $R_s$ and effective surface temperature $T(R_s) = T_{\rm DS}$, (iii) evaluate surface luminosity $L_{\rm surf}$ through Eq.~\eqref{eq:surface_lum} and DM heating $L_{\rm DM}$ via Eq.~\eqref{eq:L_dm}, and (iv) iterate the trial radius until equilibrium condition Eq.~\eqref{eq:theq} holds.
Repeating this procedure following the DS accretion, which we describe in Eq.~\eqref{eq:accretion_ds} below, yields the luminosity as a function of the DS mass and DM parameters. Over the mass range relevant for supermassive DSs, the equilibrium luminosity is well captured by the power law fit~\cite{Schwemberger:2024rvu}, and we adopt
\begin{equation}\label{eq:lum_approx}
L_{\mathrm{DS}}(M_{\mathrm{DS}}) \simeq
2.1 \times 10^{4} 
\left(\frac{M_{\mathrm{DS}}}{M_\odot}\right)^{0.85}
\left(\frac{\langle \sigma v \rangle}{10^{-26}~ \mathrm{cm^3~s^{-1}}}\right)^{0.45}
\left(\frac{m_\chi(1-f_\nu)}{100~ \mathrm{GeV}}\right)^{-0.46}
 L_\odot   ,
\end{equation}
where $M_\odot$ denotes solar mass and $L_\odot$ denotes solar luminosity.
We employ this throughout our work as baseline DS luminosity model.

\subsection{Dark matter annihilation in contracted halo}

Inside the DS  the electromagnetic and hadronic products of DM annihilation are rapidly thermalized in the optically thick interior. As a result, the annihilation power of Eq.~\eqref{eq:L_dm} does not give rise to an observable high-energy signal. Only the neutrinos at energy fraction $f_\nu$ escape. The interior luminosity therefore supports the stellar structure but does not contribute to the diffuse photon background discussed below. Outside the photosphere the situation is different. The same adiabatic contraction that enhances the DM density in the core also produces an extended overdensity in the region immediately outside the star, where the baryon density drops and the medium becomes optically thin. In this outer region  DM annihilation products can escape freely  generating an electromagnetic signal. 

\begin{figure}[t]
    \centering
    \includegraphics[width=0.5\linewidth]{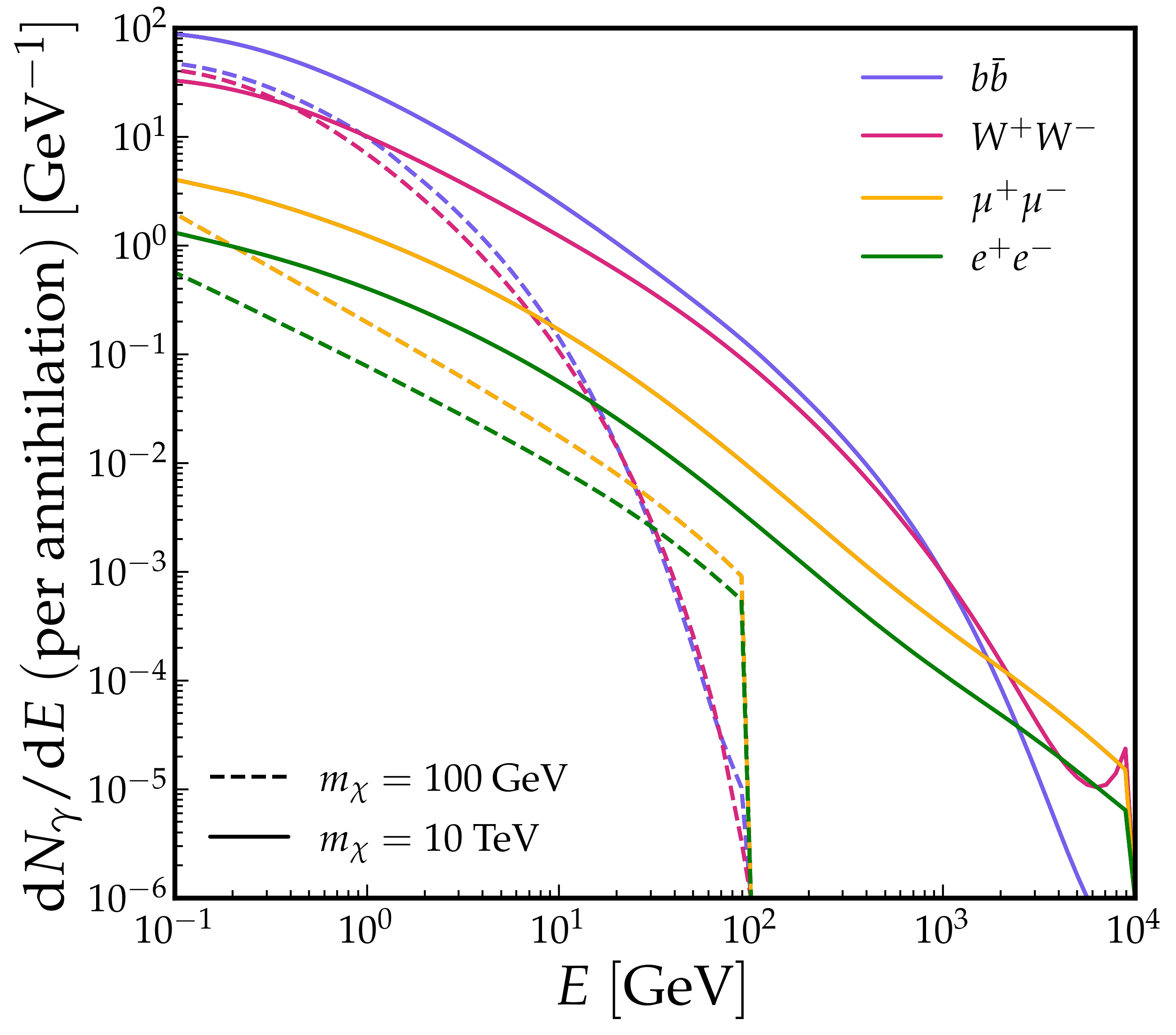}
    \caption{Photon spectra per DM annihilation for the channels $b\bar b$, $W^+W^-$, $\mu^+\mu^-$, and $e^+e^-$. Dashed curves correspond to a DM mass $m_\chi = 100~\mathrm{GeV}$, while solid curves correspond to $m_\chi = 10~\mathrm{TeV}$.  For $m_\chi\gg m_W$  internal bremsstrahlung leads to a final state radiation enhancement near the kinematic endpoint in the $W^+W^-$ channel~\cite{Bergstrom:2008gr}. Computed using data of Ref.~\cite{Cirelli:2010xx}.}
    \label{fig:spectra}
\end{figure}

The corresponding photon luminosity accounting for DM annihilations in contracted DM halo external to the DS surface is
\begin{equation}\label{eq:L_out}
L_{\rm out} =
f_\gamma \int_{R_s}^{r_h} dr  4\pi r^2 
\frac{\langle\sigma v\rangle}{m_\chi}\rho_\chi(r)^2  ,
\end{equation}
where $R_s$ is the DS photosphere radius, $r_h$ is the halo radius\footnote{For typical parameters we find that the integral of Eq.~\eqref{eq:L_out} rapidly saturates, with contributions beyond $\sim100 R_s$ affecting the result only marginally.}, and $\rho_\chi(r)$ is the contracted DM profile given in Eq.~\eqref{eq:dm_dens}.
The factor $f_\gamma$ denotes the fraction of the annihilation energy that ends up in photons. It is obtained from the photon spectrum per annihilation analogously to neutrinos as
\begin{equation}
    f_\gamma = \frac{1}{2 m_\chi} \int dE E \frac{dN_\gamma}{dE} ~. 
\end{equation}
In Fig.~\ref{fig:spectra} we show the photon spectra for representative DM masses 
$m_\chi = 100~\mathrm{GeV}$ and $m_\chi = 10~\mathrm{TeV}$ for the 
$b\bar{b}$, $W^+W^-$, $\mu^+\mu^-$ and $e^+e^-$ annihilation channels considered in this work and normalized to the total number of photons emitted per annihilation computed using data of Ref.~\cite{Cirelli:2010xx}. Increasing $m_\chi$ hardens the spectrum and modifies the photon energy fraction $f_\gamma$ entering Eq.~\eqref{eq:L_out}. These spectra are used both to determine the photon energy fraction $f_\gamma$ and subsequently to compute the diffuse flux. 
In Tab.~\ref{tab:fgamma} we list the resulting photon fractions $f_\gamma$ and the average photon energies $\langle E\rangle = \int dE~ E~ (dN_\g/dE)$ for the annihilation channels used in our analysis considering representative DM masses of $m_{\chi} = 100$~GeV and 10 TeV.

\begin{table}[t]
\centering
\renewcommand{\arraystretch}{1.0} 
\begin{tabular}{c|c|c|c|c}
\hline\hline
Annihilation&
\multicolumn{2}{c|}{$m_\chi = 100~\mathrm{GeV}$} &
\multicolumn{2}{c}{$m_\chi = 10~\mathrm{TeV}$} \\
channel & $f_\gamma$ & $\langle E\rangle$ (GeV) & $f_\gamma$ & $\langle E\rangle$ (GeV) \\
\hline
$b\bar{b}$   & 0.272 & 1.47 & 0.273 & 32.6 \\
$W^+W^-$     & 0.201 & 1.41 & 0.245 & 61.4 \\
$\mu^+\mu^-$ & 0.061 & 6.00 & 0.111 & 202 \\
$e^+e^-$     & 0.034 & 9.75 & 0.042 & 231 \\
\hline\hline
\end{tabular}
\caption{Photon energy fraction $f_\gamma$ and average energy $\langle E\rangle$ for different DM annihilation channels, considering $m_{\chi} = 100$~GeV and 10 TeV.}
\label{tab:fgamma}
\end{table}

\subsection{Dark matter annihilation in post-collapse black hole spikes}\label{sec:BH_profile}

Since DSs can grow to supermassive scales their collapse into BHs can readily provide a direct pathway to form the heavy seeds for SMBHs. The onset of gravitational Feynman-Chandrasekhar instability has been recently  examined in detail in Ref.~\cite{Freese:2025dmo}, which found that collapse typically occurs at $M_{\rm DS}~\sim~10^{6}~M_\odot$ with the precise value depending on the DM parameters and the DS evolution history. In what follows we track the evolution of these objects through collapse, after which DM again can significantly contribute to emission through annihilations in overdense DM spikes that surround newly born BHs. This phase contributes an additional diffuse photon background component extending from the time of collapse down to the present epoch.

As an initial condition we consider that the DM halo follows Navarro-Frenk-White (NFW) density profile~\cite{Navarro:1996gj}. Although we refer to this as the halo profile, we note that our formalism can readily be extended to alternative choices. The density distribution is
\begin{equation}
    \rho_h(r) =
    \frac{\rho_0}{\Big(\dfrac{r}{r_s}\Big)\Big(1+\dfrac{r}{r_s}\Big)^2} ,
\end{equation}
where $\rho_0$ is the characteristic density and $r_s$ is the scale radius. The halo radius $r_h$ is defined as the radius enclosing a mean density equal to $\Delta$ times the critical density of the Universe $\rho_c(z) = 3 H^2(z)/(8\pi G)$ at redshift $z$. Here, Hubble parameter is\footnote{Radiation contribution is neglected, as it is subdominant in the redshift range of interest.} $H(z)=H_0\sqrt{\Omega_m(1+z)^3+\Omega_\Lambda}$ and
we adopt the Planck 2018 best fit $\Lambda$CDM cosmology values of
$H_0=67.4~{\rm km~s^{-1}~Mpc^{-1}}$,
$\Omega_m=0.315$ 
and $\Omega_\Lambda=0.685$~\cite{Planck:2018vyg}.
The corresponding halo mass is
\begin{equation}
M_h = \frac{4\pi}{3} \Delta \rho_c(z)  r_h^3,
\end{equation}
and throughout we adopt the conventional choice $\Delta=200$~\cite{Prada:2011jf}.
The scale radius $r_s$ is related to the halo radius through the concentration parameter, $C=r_h/r_s$. We take $C=3.5$ for simplicity~\cite{Prada:2011jf}, which fixes $r_s=r_h/C$. With these definitions the normalization of the NFW profile becomes
\begin{equation}
\rho_0
= \rho_c(z)  
\frac{\Delta C^3}{3\Big(\ln(1+C)- \dfrac{C}{1+C}\Big)} .
\end{equation}

To determine the DM spike profile adiabatic growth of the central BH in DM halo is often assumed, leading to $\rho \propto r^{-7/3}$~\cite{Gondolo:1999ef,Bertone:2005xz}. 
In our scenario the BHs form from the rapid collapse of supermassive DSs that have already resided inside pre-existing, gravitationally contracted halos. Once the BH forms  it dominates the gravitational potential in its vicinity and can induce a spike-like enhancement in the surrounding DM distribution.
For comparison, as is typically considered in the context of primordial BHs, when a compact object grows inside an initially nearly uniform background through self-similar secondary infall~\cite{Bertschinger:1985pd,Ricotti:2007au} one expects a  $\rho\propto r^{-9/4}$  profile\footnote{Recently, it was shown that lensing of such DM halos engulfing primordial BHs can play key role in distinguishing scenarios of mixed DM~\cite{Oguri:2022fir,GilChoi:2023ahp} and can also significantly affect astrophysical environments such as   by heating of interstellar gas~\cite{Kim:2025gck}.}.

To describe the enhancement of the DM density produced by the formed BH after DS collapse we adopt the analytical parametrization of a DM spike following Ref.~\cite{Gondolo:1999ef}. The DM spike radius is taken to be a fixed fraction of the BH’s gravitational influence scale~\cite{Merritt:2003qc} 
\begin{equation}
    r_{\rm sp} \simeq 0.2  \frac{G M_{\rm BH}}{\sigma_v^2}
    \simeq 2.2~{\rm pc} 
    \left( \frac{M_{\rm BH}}{10^8 M_\odot} \right)
    \left( \frac{\sigma_v}{200~{\rm km~s^{-1}}} \right)^{-2} ,
\end{equation}
where $\sigma_v$ is the one  dimensional velocity dispersion of the surrounding matter at radii well outside the spike. We relate $\sigma_v$ to the BH mass using the empirical $M_{\rm BH}$-$\sigma_v$ relation  $\sigma_v \propto M_{\rm BH}^{1/5.64}$~\cite{McConnell:2012hz}.

\begin{figure}[t]
    \centering
    \includegraphics[width=0.5\linewidth]{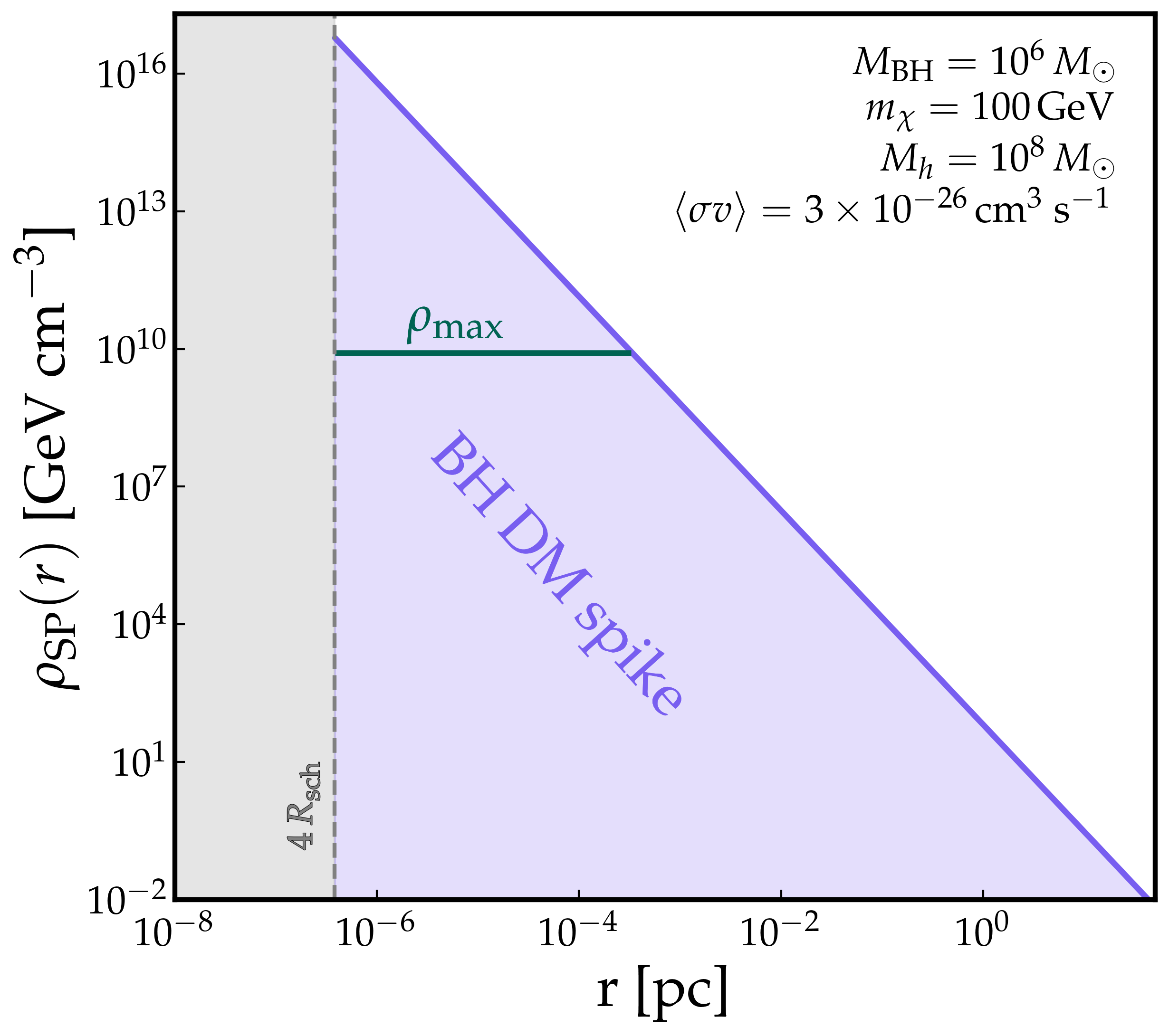}
    \caption{DM density profile of the spike surrounding a BH of mass $M_{\rm BH}=10^{6} M_\odot$ embedded in a $M_h = 10^{8} M_\odot$ halo. The solid curve denotes the power-law spike $\rho_{\rm sp}\propto r^{-\gamma_{\rm sp}}$, while the green horizontal line marks the annihilation plateau at $\rho_{\rm max}=m_\chi/(\langle\sigma v\rangle,t_{\rm BH})$, where $t_{\rm BH}\simeq1.3\times10^{10}~$yr is the time elapsed since collapse at $z_{\rm lim}=15$.}
    \label{fig:density_spike}
\end{figure}

The spike density profile is then
\begin{equation} \label{eq:spike_density}
    \rho_{\rm sp}(r) =
    \rho_h(r_{\rm sp})
    \left( \frac{r}{r_{\rm sp}} \right)^{-\gamma_{\rm sp}} ,
\end{equation}
where the spike slope can be written in terms of adiabatic contraction relation $\gamma_{\rm sp} = (9- 2\gamma)/(4-\gamma)$, which can denote more peaked $\gamma > 0$ or cored $\gamma \simeq 0$ profiles. 
As our reference value, we adopt a fixed inner slope $\gamma_{\rm sp} = 7/3$ that corresponds to the spike index obtained for a BH embedded in a pre-existing NFW-like halo~\cite{Gondolo:1999ef}. For $r < 4 R_{\mathrm{sch}}$,
where $R_{\mathrm{sch}} = 2GM_{\mathrm{BH}}$ is the Schwarzschild radius of BH of mass $M_{\rm BH}$, the DM particles are captured by BH. Relativistic corrections can further modify this to $2R_{\rm sch}$~\cite{Sadeghian:2013laa}. Here, we conservatively consider that spike vanishes for $r < 4R_{\mathrm{sch}}$.
In Fig.~\ref{fig:density_spike} we show the DM spike density profile outside the BH for $M_{\mathrm{BH}}=10^6 M_{\odot}$ in a halo of mass $M_h = 10^8  M_\odot$.

After formation, the DM spike can undergo complex evolution.
For DM spikes existing for a significant fraction of the Hubble time DM annihilations in the innermost regions can progressively deplete the DM density. A convenient way to to capture this effect is through the maximal density description \cite{Gondolo:1999ef} 
$\rho_{\mathrm{max}} \simeq  m_{\chi}/(\langle \sigma v\rangle  t_{\mathrm{BH}})$,
where $t_{\rm BH}$ is the time during which the spike has been subject to annihilations. In our scenario the BH forms at $z_{\rm lim}=15$, so $ t_{\mathrm{BH}} \simeq t_0- t(z_{\rm lim}) \simeq 1.3 \times 10^{10}~\mathrm{yr}$, corresponding to the time interval between collapse and today. This effectively assumes that the inner spike has been limited by DM annihilation  throughout its post-collapse evolution. In principle, one could evolve $\rho_{\rm max}$ with redshift by promoting $t_{\rm BH}\to \Delta t(z)=t(z)-t(z_{\rm lim})$. However, a fully consistent and realistic treatment would need to incorporate additional effects including gravitational heating by stars~\cite{Ahn:2007ty,Ullio:2001fb,Merritt:2003qk,Bertone:2005hw,Bertone:2005xv,Balaji:2023hmy}, halo and BH mergers, deviations from spherical symmetry and centrophilic orbits that are particularly relevant in triaxial halos where supermassive DS can form~\cite{Freese:2010aw}, accretion as well as possible spike regrowth. Modeling of these processes, as well as effects of BH evolution, requires detailed simulations and lies beyond the scope of this work.

Given these uncertainties we consider three limiting prescriptions for the post-collapse DM spike in our analysis. First, we adopt a conservative scenario in which the spike is truncated at $\rho_{\rm max}$ using the full time interval from $z_{\rm lim}$ to $z=0$, leading to a strongly depleted inner density. Second, we examine an optimistic case in which no cutoff is applied and the spike retains its original power-law form without any evolution effects. Finally, we include a maximally conservative scenario in which a spike either fails to form or is completely disrupted on the relevant timescales, producing no appreciable contributions to the photon background. A realistic spike evolution is thus expected to lie between these scenarios.

\begin{figure}[t]
    \centering
    \begin{subfigure}[t]{0.49\textwidth}
        \centering
        \includegraphics[width=\linewidth]{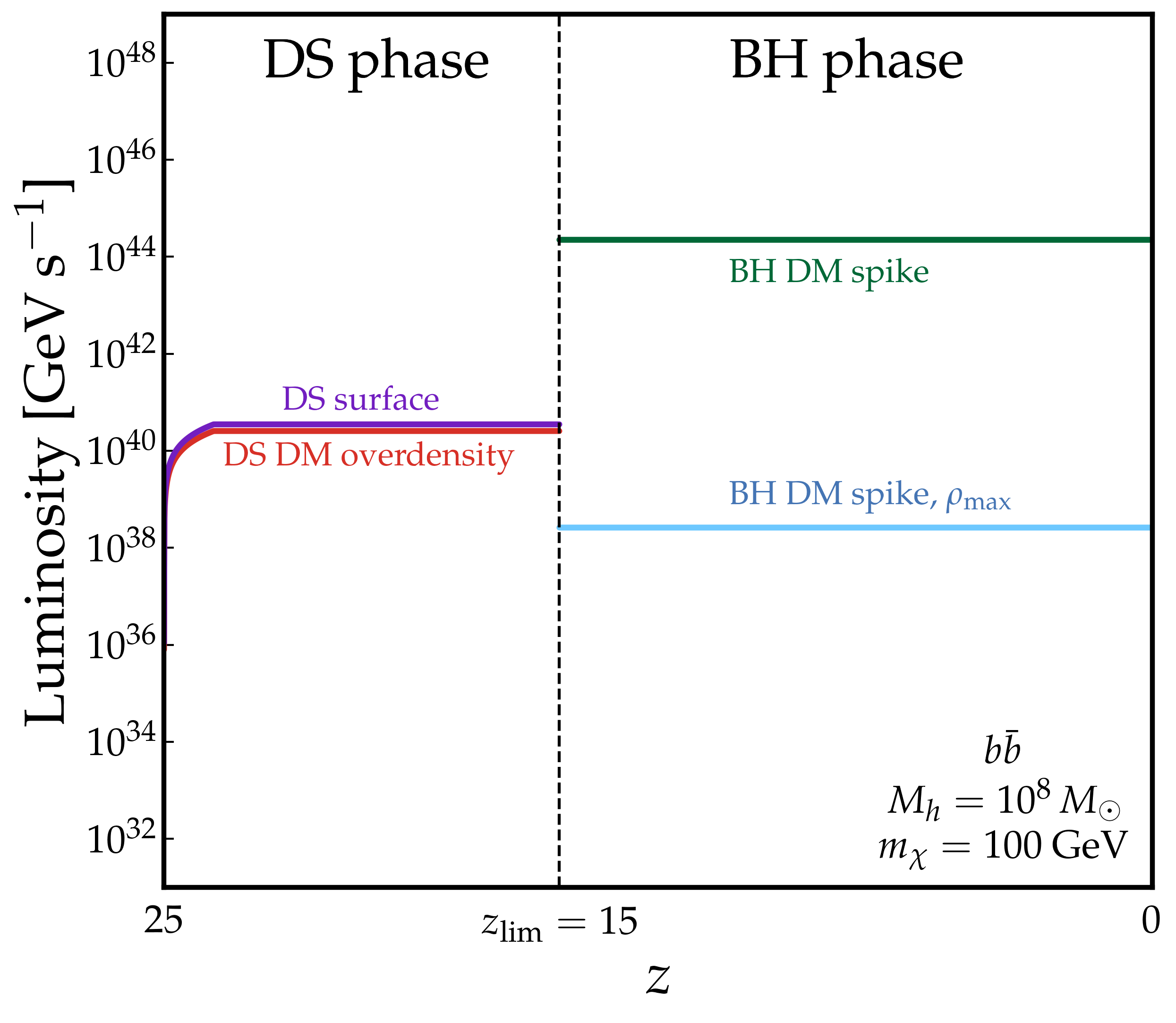}
    \end{subfigure}
    \hspace{0.1em}
    \hfill
    \begin{subfigure}[t]{0.49\textwidth}
        \centering
        \includegraphics[width=\linewidth]{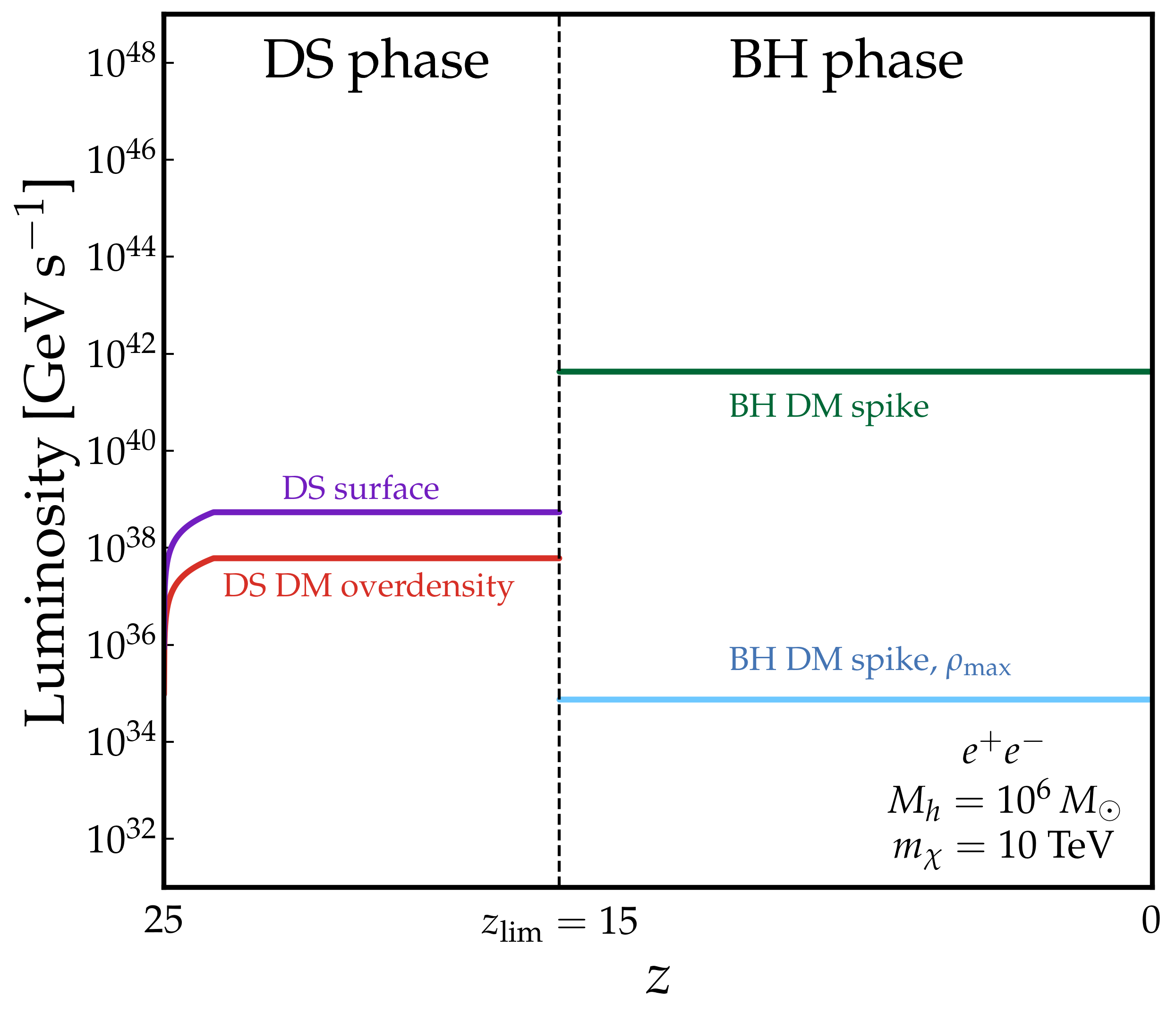}
    \end{subfigure}
\caption{
Luminosity evolution of a DS forming at $z_{\rm form}=25$ and collapsing at $z_{\lim}=15$.
\textbf{Left:} Benchmark model with halo mass $M_h = 10^8 M_\odot$, DM mass $m_\chi = 100$~GeV, and annihilation channel $\chi\chi \to b\bar{b}$, assuming a thermal relic cross section $\langle\sigma v\rangle = 3\times10^{-26} {\rm cm}^3 {\rm s}^{-1}$. 
The DS surface luminosity during its accreting phase (purple), the luminosity from DM annihilation in the DS-induced contracted halo (red), and the post-collapse luminosity from the BH DM spike (green) are shown, together with a conservative possible limit imposed by $\rho_{\max}$ (light blue). 
\textbf{Right:} Same as the left panel, but for $M_h = 10^6 M_\odot$, $m_\chi = 10$~TeV, considering leptonic DM annihilation channel $\chi\chi \to e^+e^-$. 
}
    \label{fig:lum_comparison}
\end{figure}

The photon luminosity from DM annihilation from a spike surrounding a BH of mass $M_{\rm BH}$ is
\begin{equation}\label{eq:lum_BH}
    L_{\mathrm{BH}} \simeq  f_\gamma\int_{4R_{\mathrm{sch}}}^{r_h} dr   4 \pi r^2 \frac{\langle \sigma v \rangle \rho_{\mathrm{sp}}^2}{m_\chi}~.
\end{equation} 
In Fig.~\ref{fig:lum_comparison} we illustrate the luminosity of a DS forming at $z_{\rm form}=25$ with an initial mass
 $M_{{\rm DS},i}=5~M_\odot$ and growing through accretion, considering two benchmark scenarios. 
We model the DS mass growth as
\begin{equation} \label{eq:accretion_ds}
    M_{\rm DS}(t) = M_{{\rm DS},i} + \dot{M}_{\rm DS} t ,
\qquad
\dot{M}_{\rm DS} = 10^{-9}~M_h~{\rm yr}^{-1} ,
\end{equation}
which captures a sustained accretion phase enabled by the low effective temperature and extended radius of DSs calculated in Sec.~\ref{sec:polytrope}. 
For the halo masses considered in this work the resulting accretion rates span from $10^{-3}$ to $1~M_\odot \rm yr^{-1}$. 
These rates are consistent with those of Ref.~\cite{Rindler-Daller:2014uja}. Although we include halos as massive as $10^9~M_\odot$, as we discuss below in Eq.~\eqref{eq:halo_masses}, such objects are extremely rare and we find that they do not significantly alter the overall luminosity.

For a given host halo mass, we impose the bound
\begin{equation}\label{eq:mass_lim}
M_{\rm DS}^{\rm lim} = 10^{-2} M_h,
\end{equation}
ensuring that DSs remain a minor component of the halo mass while still allowing growth to supermassive scales. For the accretion rates in Eq.~\eqref{eq:accretion_ds}, this limit is typically reached well before $z_{\rm lim}=15$. As a result, the integrated luminosity is only weakly sensitive to $\dot M_{\rm DS}$, since the star spends most of its lifetime in regime where its mass is $M_{\rm DS}^{\rm lim}$. Since $L_{\rm DS}\propto M_{\rm DS}^{0.85}$, reducing $M_{\rm DS}^{\rm lim}$ by an order of magnitude decreases the luminosity by a factor of $\sim7$.

In Fig.~\ref{fig:lum_comparison} we illustrate the distinct DS emission contributions, considering two different scenarios of $b\bar{b}$, $M_h = 10^8 M_{\odot}$, $m_{\chi} = 100$~GeV and $e^+e^-$, $M_h = 10^6 M_{\odot}$, $m_{\chi} = 10$~TeV
assuming a thermal relic annihilation cross-section $\langle\sigma v\rangle = 3\times10^{-26}~{\rm cm^{3}~s^{-1}}$~\cite{Steigman:2012nb}
.
At $z_{\rm lim}=15$ we assume the DS collapses directly into a BH. The stellar surface luminosity then vanishes, leaving only the emission from annihilations in the BH spike. We show both limiting prescriptions for this luminosity with and without the $\rho_{\rm max}$ cutoff. Considering $e^+e^-$ annihilation channel instead of $b\bar{b}$ primarily affects the photon yield $f_\gamma$ (see Tab.~\ref{tab:fgamma}), which leads to observed suppression of DS-induced contracted DM halo luminosity compared to DS surface luminosity that is not strongly sensitive to the DM annihilation channel. Channels with smaller $f_\gamma$, such as $e^+e^-$, suppress the emission that scales directly with the photon energy fraction.
Reducing the halo mass decreases all luminosity components since the DS reaches a smaller maximum mass, as set by Eq.~\eqref{eq:mass_lim}. Increasing the DM mass $m_\chi$ generally decreases the luminosity as well. The exception is the ``BH DM spike, $\rho_{\rm max}$'' case, where the non-trivial dependence of $\rho_{\rm max}$ on $m_\chi$ partially compensates this suppression by allowing for larger maximum density.

\section{Dark Star Population} \label{sec:ds_population}

To determine the total diffuse background produced by DSs we need to account for their entire cosmological population. Following the diffuse neutrino background analysis of Ref.~\cite{Schwemberger:2024rvu}  we compute the population of DSs that can reproduce the abundance of SMBH seeds. Similar to Pop~III stars, DSs are considered to form in the centers of DM mini-halos with masses $10^{5}$-$10^{7}~M_\odot$ at redshifts $z\sim20$~\cite{Abel:2001pr,Bromm:2001bi,Yoshida:2006bz}. These regions exhibit high baryonic and DM densities and adiabatic contraction steepens the local DM profile  boosting annihilation rates and providing a sustained heating source for the protostellar core. We model DS formation as tracing the halo assembly history, with a fraction of newly formed halos producing DSs instead of Pop~III stars when heating from DM annihilation becomes comparable to or exceeds the energy input from nuclear fusion.

The comoving halo number density $n_h(M_h,z)$ is described by the mass function
\begin{equation}\label{eq:haloformationrate}
\frac{dn_h}{dM_h} = \frac{\rho_m}{M_h}
\sqrt{\frac{2A^2\beta}{\pi}} 
\Big(1+(\beta \nu^2)^{-p}\Big) 
e^{-\beta \nu^2/2} 
\frac{d\nu}{dM_h} ,
\end{equation}
where $\rho_m$ is the present day mean matter density~\cite{Planck:2018vyg},
$\nu = \delta_c/(D(z) S(M_h))$,
$\delta_c$ is the critical linear overdensity for collapse~\cite{1992ARA&A..30..499C},
$D(z)$ is the linear growth factor~\cite{growth_factor} normalized to $D(0)=1$,
and $S(M_h)$ is the variance of the smoothed linear density field.
The choice $(A,\beta,p)=(0.5,1,0)$ reproduces the Press-Schechter mass function~\cite{Press:1973iz}. In App.~\ref{app:ST} we compare this to the Sheth-Tormen mass function described by $(A,\beta,p)=(0.322,0.707,0.3)$~\cite{Sheth:1999mn}.
The variance of the initial density fluctuation field is
\begin{equation}
S^2(M_h) = \frac{1}{2\pi^2}
\int W^2(kR_{M_h}) P_\delta(k) k^2 dk~,
\end{equation}
where $W(x)=3(\sin x- x\cos x)/x^3$ is the real space top-hat window function\footnote{Our analysis can be readily extended to other possibilities.} smoothing fluctuations on the scale
$R_{M_h}=(3M_h/4\pi\rho_m)^{1/3}$, and $P_\delta(k)$ is the linear matter power spectrum~\cite{Pierpaoli:2000ip,Bardeen:1985tr}.

We can then directly determine the population of DSs by considering that each formed halo of mass $M_h$ hosts a DS with probability $f_{\rm SMDS}$ within the redshift range of interest.
We take $d^2n_h/(dt dM_h)$ to denote the halo formation rate. The comoving number density of DSs for host halo mass $M_h$ present at redshift $z$ and with age $t'$ can be modeled as
\begin{equation}\label{eq:n_SMDS}
\frac{dn_{\rm DS}}{dM_h}(M_h,z;t')
= f_{\rm SMDS}
\int_{z(t)}^{z_{\rm form}} dz' 
\left|\frac{dt}{dz'}\right| 
\frac{d^2n_h}{dt dM_h} \left(M_h,z'(t-t')\right),
\end{equation}
where $t = t(z)$ is the cosmic time at redshift $z$. The time-redshift relation is $ 
dt/dz = 1/(H(z)(1+z))$. 
 Here, we can obtain $n_{\rm SMDS}$ by integrating Eq.~\eqref{eq:n_SMDS} over the range of halos, as given in Eq.~\eqref{eq:halo_masses} below.

Motivated by the connection of supermassive DSs to SMBH seeds and by the expectation that they form only during the earliest stages of structure assembly  we restrict DS formation to the epochs in which minihalos first collapse and gas cooling becomes efficient. In our baseline scenario we consider that DS formation operates during\footnote{The specific value of the considered $z_{\rm form}=25$ is not essential, since the halo formation rate is strongly suppressed at higher redshifts. Thus, extending this redshift interval does not significantly affect our results.} $z_{\rm lim}=15 < z < z_{\rm form}=25$, corresponding to DS lifetimes of order $\mathcal{O}(10^8)$ yrs. Below $z_{\rm lim}$ we impose a sharp cutoff on the DS survival, consistent with the observed high-redshift galaxy population and with the expectation that DSs surviving at redshifts significantly below $z\sim 15$ could be potentially visible in data surveys such as JWST~\cite{Freese:2010re}. We discuss more gradual transition and its impact on the DS population below  Eq.~\eqref{survival}.

Within the redshift window between $z_{\rm form}$ and $z_{\rm lim}$ each newly formed halo of mass $M_h$ is considered to host a DS with probability $f_{\rm SMDS}$. We can then determine $f_{\rm SMDS}$ by requiring that the comoving number density of supermassive DSs matches that required to seed the SMBH population, which can be stated as the following approximate criteria~\cite{Munoz:2021sad}
\begin{equation} \label{eq:smbhds}
     \dfrac{n_{\rm SMDS}(z)}{(1+z)^3} \sim \dfrac{\rho_B}{10^{10}~M_\odot}~.
\end{equation}
Using $\rho_B \simeq 4\times 10^{-31} ~\mathrm{g~cm^{-3}}$~\cite{Planck:2018vyg}, condition of Eq.~\eqref{eq:smbhds} is approximately satisfied for a constant halo occupation fraction $f_{\rm SMDS} \simeq 10^{-2}$. The total DS luminosity scales linearly with $f_{\rm SMDS}$. Consequently, adopting an alternative $f_{\rm SMDS}=10^{-3}$ uniformly rescales all fluxes by an additional factor of $10^{-1}$ without altering their spectral shape or qualitative behavior.

In what follows, we restrict our attention to minihalos in the interval
\begin{equation}\label{eq:halo_masses}
M_{h,\min} \sim 10^6 M_\odot,
\qquad
M_{h,\max} \sim 10^9 M_\odot .
\end{equation}
Here, the lower limit reflects the requirement for molecular-hydrogen cooling~\cite{Ilie:2011da}, while the upper limit captures the relevant population. More massive halos are exceedingly rare in the redshift range of interest. Varying $M_{h,\min}$ alters the abundance of potential hosts, but imposing consistency with the SMBH seeding implies adjusting $f_{\rm SMDS}$ accordingly, keeping the resulting impact modest.

\section{Diffuse Electromagnetic Background} \label{sec:flux}

 We have derived the luminosities associated with all stages of DS evolution: the DM-powered stellar emission, the radiation produced in the contracted DM overdensity surrounding the DS  and the annihilation signals emerging from the DM spike formed around the BH remnant after collapse. We now combine these single source luminosities and discuss their cosmological implications by computing the diffuse photon background associated with the full population of DSs and their remnants.
This requires combining the luminosity of each source with the formation history and abundance of the DS population derived above and propagating the emitted photons to the present epoch while accounting for cosmological redshifting as well as attenuation effects.

\subsection{Stellar surface emission}

DSs emit a thermal component associated with radiation from their photospheres. Whereas the peak of the DM annihilation-induced signal is directly controlled by the DM mass, typically considered of the GeV-scale, the thermal emission component emerges at much lower energies of order eV reflecting the effective surface temperatures of DSs. 

Inside the DS the annihilation energy is almost entirely trapped and thermalized in the optically thick interior thus providing the dominant source of pressure support. At the photosphere radiation can escape and gives rise to the thermal emission.
The structural properties of DSs, including the photospheric temperature $T_{\rm DS}$ and radius $R_{\rm DS}$, can be obtained by solving the stellar equilibrium equations within the polytropic model introduced in Sec.~\ref{sec:polytrope}, where the internal DM heating rate provides the main energy source balancing gravity for negligible nuclear fusion. 
Considering the DS photosphere radiates approximately as a blackbody, the normalized photon spectrum at emission is
\begin{equation} \label{eq:thermal_spectra}
    \frac{dN^{\rm th}_\gamma}{dE_{\rm th}} = 
    A   \frac{(E_{\rm th}/k_B T_{\rm DS})^2}{e^{E_{\rm th}/(k_B T_{\rm DS})}- 1}~.
\end{equation}
Here,  $A$ is set to normalize the distribution such that $\int_0^\infty dE_{\rm th} (dN^{\rm th}_\g/dE_{\rm th})=1$, and $\langle E_{\rm th}\rangle_{\rm th} = \int dE_{\rm th}~ E_{\rm th}~ (dN^{\rm th}_\g/dE_{\rm th})$. Typical values for supermassive DSs in our scenario are $T_{\rm DS} \sim 10^4~{\rm K}$, corresponding to photon energies $E \sim {\rm eV}$.

\begin{figure}[t]
    \centering
    \includegraphics[width=0.5\linewidth]{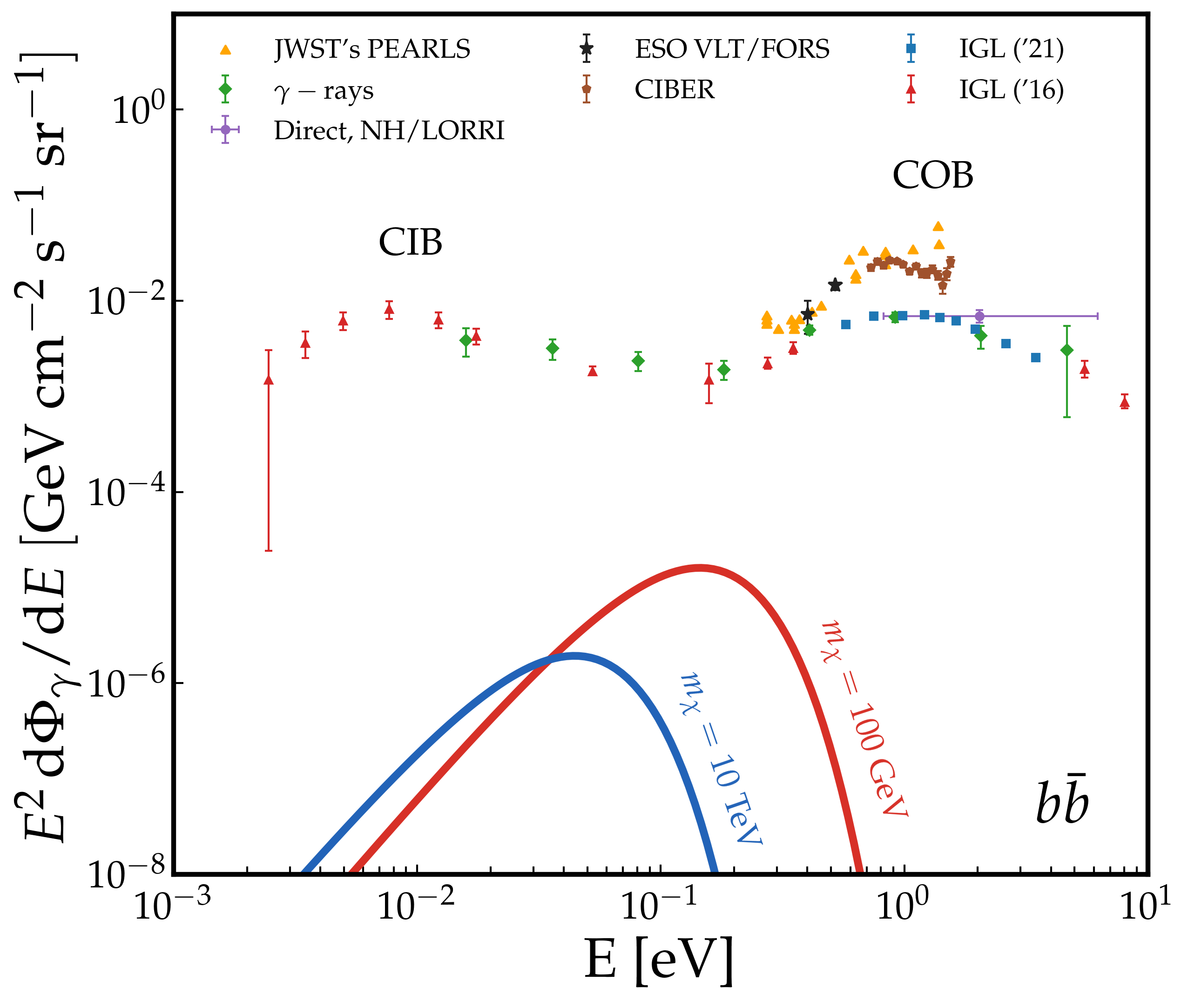}
\caption{
Predicted diffuse thermal fluxes from a cosmological population of supermassive DSs, shown for representative DM masses $m_\chi=100$~GeV and $m_\chi=10$~TeV, assuming the benchmark annihilation channel $\chi\chi \to  b\bar b$ and the reference relic cross-section $\langle\sigma v\rangle = 3\times10^{-26}~\mathrm{cm^3~s^{-1}}$.
Observational data from JWST/PEARLS~\cite{Windhorst:2022rrv}, $\gamma$-ray constraints~\cite{Greaux:2024wyc}, New Horizons LORRI measurements~\cite{Postman:2024erl}, ESO VLT/FORS dark cloud observations~\cite{Mattila:2017zrp}, CIBER near-infrared background data~\cite{Matsuura:2017lub}, and integrated galaxy light (IGL) estimates from~\cite{Driver:2016krv} and~\cite{Koushan2021} are shown.
The predicted supermassive DS emission peaks in the near-infrared and remains below current observational limits.} 
    \label{thermal}
\end{figure}

The corresponding single-object luminosity is given by Eq.~\eqref{eq:lum_approx}.  Averaging this emission over the DS population defined in Sec.~\ref{sec:ds_population}, we obtain the comoving luminosity density
\begin{equation}\label{eq:DS_comoving_emissivity} L_{\rm DS}^{\rm pop}(z)  =  \int_{t(z_{\rm form})}^{t(z)}   dt' \int_{M_{h,\min}}^{M_{h,\max}}   dM_h  L_{\rm DS}(M_h,t')  f_{\rm surv}(z)  \frac{d^2 n_{\rm DS}}{dM_h dt}  \left(M_h, z(t-t')\right) , \end{equation}
where $d^2 n_{\rm DS}/(dM_h,dt)$ is the DS formation rate and $t'$ is the stellar age.
Here, the survival factor 
\begin{equation} \label{survival}
f_{\rm surv}(z)=1-\Theta(z_{\rm lim}-z),
\end{equation}
ensures that DSs contribute only while they exist so that all DSs disappear at the collapse redshift $z_{\rm lim}$. 
A more gradual survival function is arguably more realistic, since DS lifetimes may vary with their local environment. 
As shown in Ref.~\cite{Schwemberger:2024rvu}, however, adopting such a smooth $f_{\rm surv}(z)$ typically leads to a reduced diffuse flux. 
This occurs because the surviving population at $z\lesssim 15$ can be constrained by JWST observations, which in turn forces a reduced abundance at earlier times. 
These considerations highlight the model dependence inherent for a particular parametric form of $f_{\rm surv}(z)$.

A cosmological population of DSs produces a diffuse thermal background obtained by integrating the individual emission over redshift and formation history. The observed differential flux is
\begin{equation}
    \frac{d \phi^{\rm th}_\gamma}{dE_{\rm th}}(E_{\rm th}) = \int_{z_{\mathrm{lim}}}^{z_{\rm form}} dz   \bigg|\frac{dt}{dz}\bigg|\frac{(1+z)}{\langle E\rangle_{\rm th}} L_{\rm DS}^{\rm pop}(z)
\frac{dN^{\rm th}_\gamma}{dE_{\rm th}}\bigg|_{E_{\rm th}'=(1+z)E_{\rm th}}e^{-\tau(E_{\rm th}, z)} ~.
    \label{eq:thermal_flux}
\end{equation}
Here, the factor $e^{-\tau(E_{\rm th},z)}$ accounts for the photon optical depth $\tau(E_{\rm th},z)$. 
In contrast to the energetic DM annihilation signals discussed below, attenuation can be safely neglected for the thermal component. The DS photons have characteristic energies of order eV and the intergalactic medium is effectively transparent to photons with energies $\lesssim 13.6~$eV~\cite{Haiman:1999mn}. Thus we set $e^{-\tau(E_{\rm th},z)}\simeq 1$ throughout this work.

In Fig.~\ref{thermal} we show the predicted diffuse thermal flux from a cosmological population of supermassive DSs for representative DM masses of $m_\chi=100$~GeV and $m_\chi=10$~TeV  in the benchmark annihilation channel $\chi\chi\to b\bar b$, together with current observational upper limits~\cite{Koushan2021,Greaux:2024wyc,Postman:2024erl,Windhorst:2022rrv,Matsuura:2017lub,Mattila:2017zrp}. The most stringent constraints are those derived from Hubble Space Telescope (HST) measurements of the optical extragalactic background light (EBL)~\cite{Driver:2016krv}.
Considering alternative annihilation channels has only a minor impact on the predicted thermal flux. This dependence enters through the luminosity expression in Eq.~\eqref{eq:lum_approx}, which scales as $(1-f_\nu)^{-0.46}$, and through the surface temperature $T_{\rm DS}$ in Eq.~\eqref{eq:thermal_spectra}. Since $T_{\rm DS}\propto L_{\rm DS}^{1/4}$ in Eq.~\eqref{eq:surface_lum}, the resulting variation in the thermal spectrum is not significant.

The resulting spectra exhibit a broad peak around the near infrared regime, reflecting the typical surface temperatures of DSs. 
As the DM particle mass increases, the photospheric temperature slightly rises due to the more compact stellar configurations required to balance the star. This leads to a modest shift of the spectral maximum toward higher energies. However, the overall normalization of the flux remains nearly unchanged, since the flux is primarily determined by $L_{\rm DS}^{\rm pop}(z)$. At the observational level, the predicted DS thermal emission lies significantly below all current limits on the cosmic optical and infrared background.

As we show in the following sections, constraints from this thermal component are significantly weaker than those obtained from the energetic DM annihilation signals originating from the DM overdensities surrounding DSs.

\subsection{Emission from halo dark matter annihilation}\label{sec:DS_flux}

We now consider the diffuse photon background arising from DM annihilation in the external, adiabatically contracted overdensity surrounding DSs. The single object luminosity is given by Eq.~\eqref{eq:L_out}, evaluated using the contracted DM profile of Eq.~\eqref{eq:dm_dens}. The respective population contribution $L_{\rm out}^{\rm pop}(z)$ is then computed analogously to Eq.~\eqref{eq:DS_comoving_emissivity}.
The diffuse flux observed today from this external annihilation component is
\begin{equation}\label{eq:DS_phase_flux} \frac{d\phi^{\rm out}_\gamma}{dE}(E) = \int_{z_{\rm lim}}^{z_{\rm form}} dz \left|\frac{dt}{dz}\right| \frac{(1+z)}{\langle E\rangle} L_{\rm out}^{\rm pop}(z) \frac{dN_\gamma}{dE}\bigg|_{E'=(1+z)E} e^{-\tau(E,z)} , \end{equation}
where $dN_\gamma/dE$ is the photon spectrum per annihilation taken from Ref.~\cite{Cirelli:2010xx}, $E'=(1+z)E$, and $\langle E\rangle$ is the channel-dependent mean photon energy listed in Table~\ref{tab:fgamma}.

Unlike the thermal surface emission, the energetic photons produced in the DS-induced external DM overdensity can experience substantial attenuation as they propagate cosmological distances. At early epochs  $\gamma$-rays are absorbed through photoionization, Compton scattering, pair production on gas  and photon-photon interactions with the ambient radiation fields including cosmic microwave background radiation (CMB), infrared background, and ultraviolet (UV), optical (EBL) backgrounds~\cite{Cirelli:2010xx}. These processes suppress the flux emitted at high energies or high redshift.

\begin{figure}[t]
    \centering
    \includegraphics[width=0.5\linewidth]{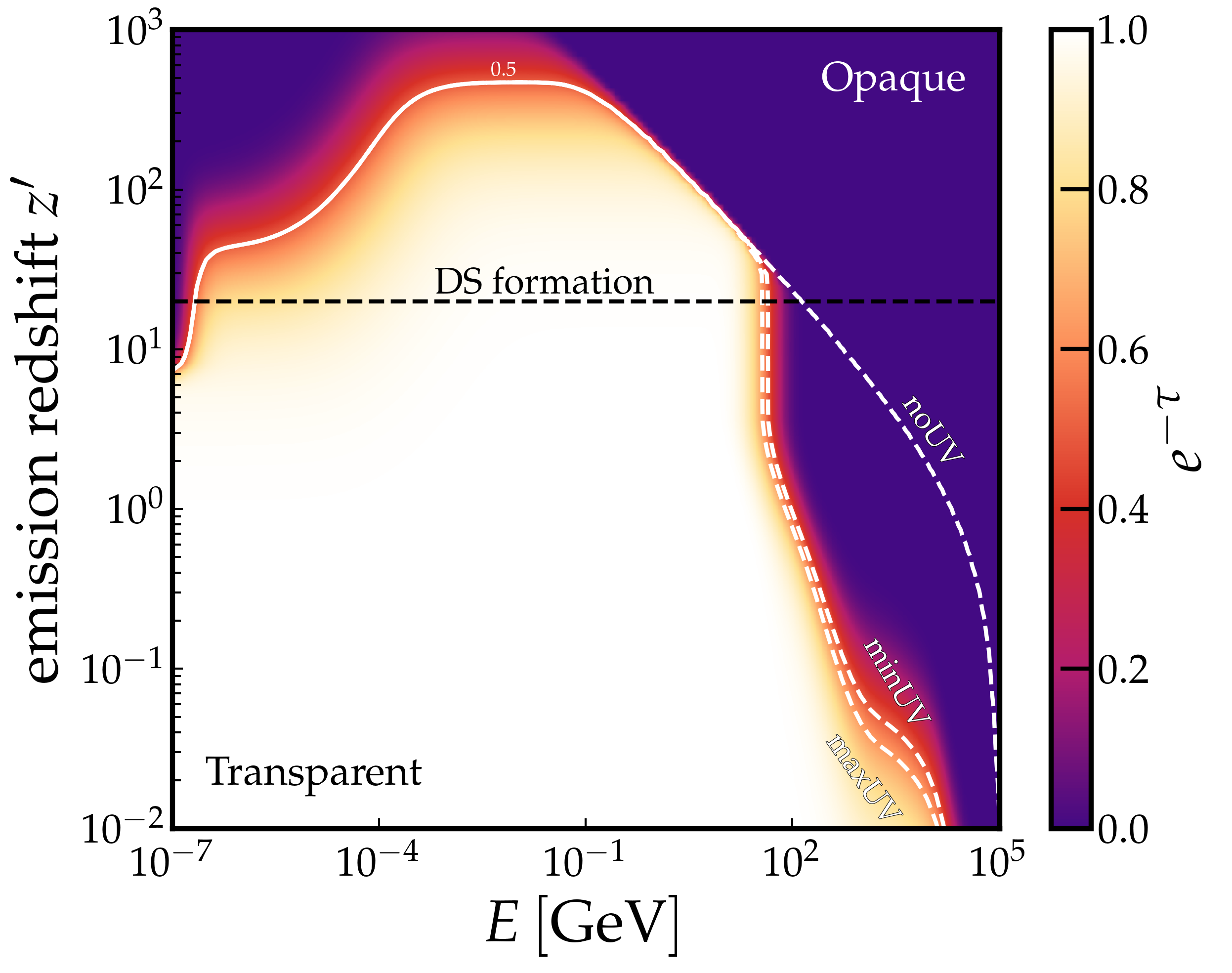}
\caption{
Optical depth $\tau(E,z')$ as a function of observed photon energy and emission redshift.
Colored regions indicate the transition from transparent to opaque propagation, with attenuation dominated by photoionization, Compton scattering, pair production on gas, and photon-photon pair production.
The dashed black line marks the characteristic formation epoch of DSs at $z_{\rm form} = 25$.
White dashed curves show the $e^{-\tau}=0.5$ contour for the noUV, minUV, and maxUV background prescriptions. In this work we adopt the minUV model, motivated by blazar observations~\cite{MAGIC:2008sib}.
}
    \label{fig:optdepth}
\end{figure}

In Fig.~\ref{fig:optdepth} we illustrate the corresponding optical depth $\tau(E,z)$ as a function of emission redshift and observed energy. The Universe becomes rapidly opaque at large $z$ and $E$, and even for formation epochs as early as $z_{\rm form}\sim25$, photons above $\sim100$~GeV are efficiently absorbed. This defines an effective transparency window that determines which portion of the intrinsic DS spectrum can reach us today. The dominant astrophysical uncertainty in computing $\tau$ arises from the UV component of the EBL. Following Ref.~\cite{Cirelli:2010xx}, we adopt the ``minUV'' model motivated by recent blazar observations~\cite{MAGIC:2008sib} as our reference choice. For completeness, we also examine two alternative prescriptions including a vanishing UV background (``noUV'') and an enhanced UV background (``maxUV''), as well as the limiting case of no attenuation. These span the plausible range of UV photon densities, and their impact is discussed in App.~\ref{app:UVmodels}.

\begin{figure}[t]
    \centering
        \includegraphics[width=0.5\linewidth]{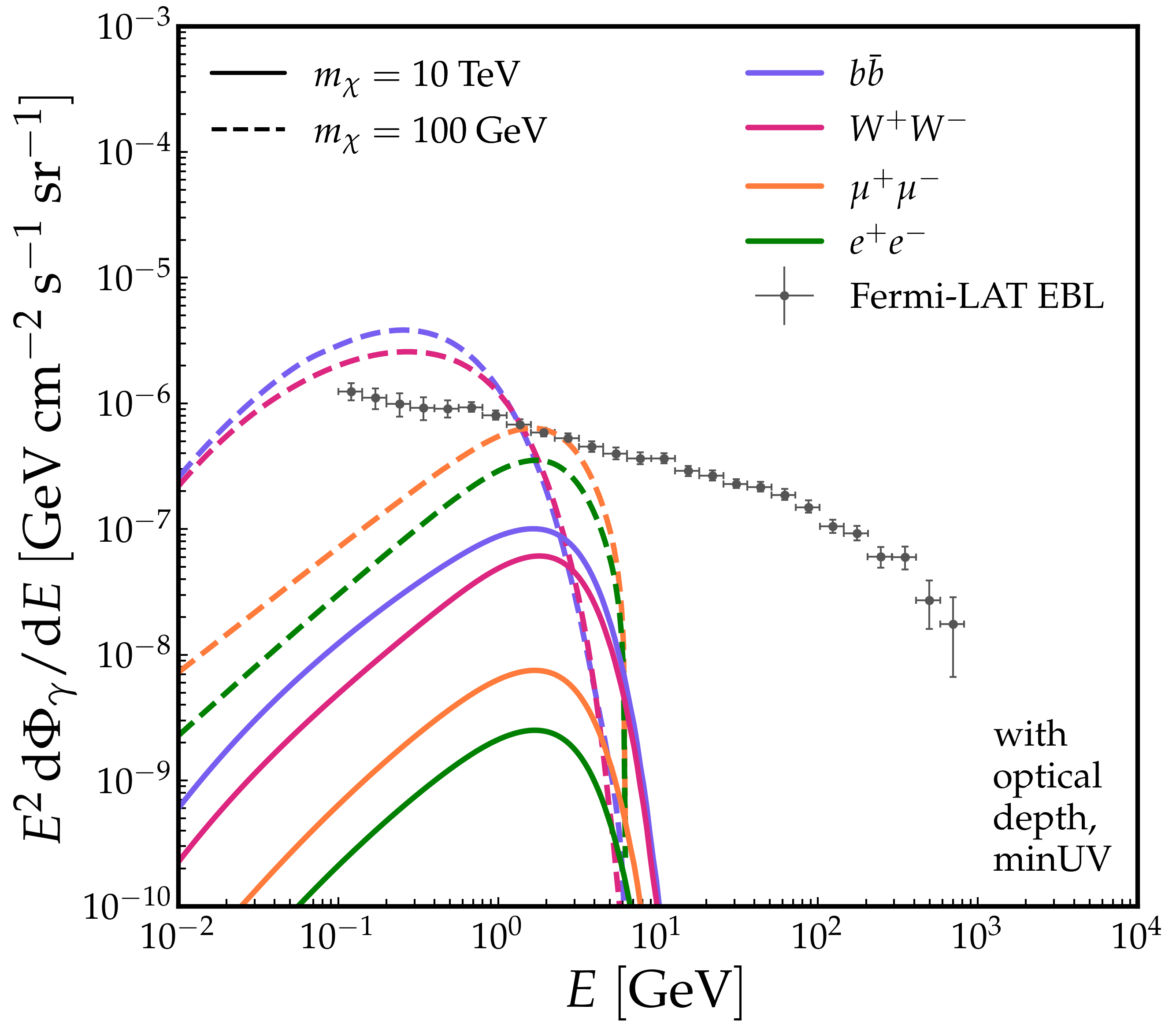}
    
\caption{
Diffuse photon flux from external DM annihilation in the DS-induced overdensity, shown for two representative DM masses and for the four annihilation channels considered. The spectral shapes reflect the interplay between the intrinsic photon spectra (see Fig.~\ref{fig:spectra}), the channel-dependent photon energy fractions $f_\gamma$ (see Tab.~\ref{tab:fgamma}), and cosmological attenuation encoded in the optical depth. All curves correspond to our reference minUV prescription for the optical depth. For comparison, we overlay the Fermi-LAT measurements of the extragalactic $\gamma$-ray background, which provide the relevant observational benchmark in this energy range.
}

    \label{fig:DS_fluxes}
\end{figure}

The resulting diffuse flux, including cosmological attenuation, is shown in Fig.~\ref{fig:DS_fluxes}. We display the external DS annihilation contribution for two representative DM masses and for the four annihilation channels considered in this work. The spectral shapes reflect the combination of the intrinsic photon spectra in Fig.~\ref{fig:spectra}, the photon energy fractions $f_\gamma$ listed in Tab.~\ref{tab:fgamma}, and attenuating effects of the optical depth.
As expected, the spectra harden for larger DM masses and for leptonic final states, while $f_\gamma$ controls the fraction of the annihilation energy emitted as photons, and therefore sets the overall normalization. The predicted fluxes are compared with the Fermi-LAT measurements of the extragalactic $\gamma$-ray background, which provide the most directly relevant observational constraints in this energy range. Results for alternative UV-background models are shown in App.~\ref{app:UVmodels}.

In App.~\ref{app:NFW} we compare the resulting flux with that expected in conventional indirect detection scenarios where the DM distribution follows a standard NFW halo profile. We find that the relative enhancement produced by the DS-induced overdensities can be substantial, corresponding to a boost factor of $f_{\rm DS}\sim \mathcal{O}(10^{5})$.

\subsection{Emission from black hole spikes}\label{sec:BH_flux}

At the end of its lifetime, each supermassive DS is expected to collapses to a massive BH, around which a dense DM spike develops. The corresponding annihilation luminosity of a single remnant is given in Eq.~\eqref{eq:lum_BH}, where the steep spike profile leads to a substantial enhancement of the annihilation rate in the innermost region. This luminosity provides the source term for the late-time diffuse background produced by DS remnants.

\begin{figure}[t]
    \centering
    \begin{subfigure}[t]{0.495\textwidth}
        \centering
        \includegraphics[width=\linewidth]{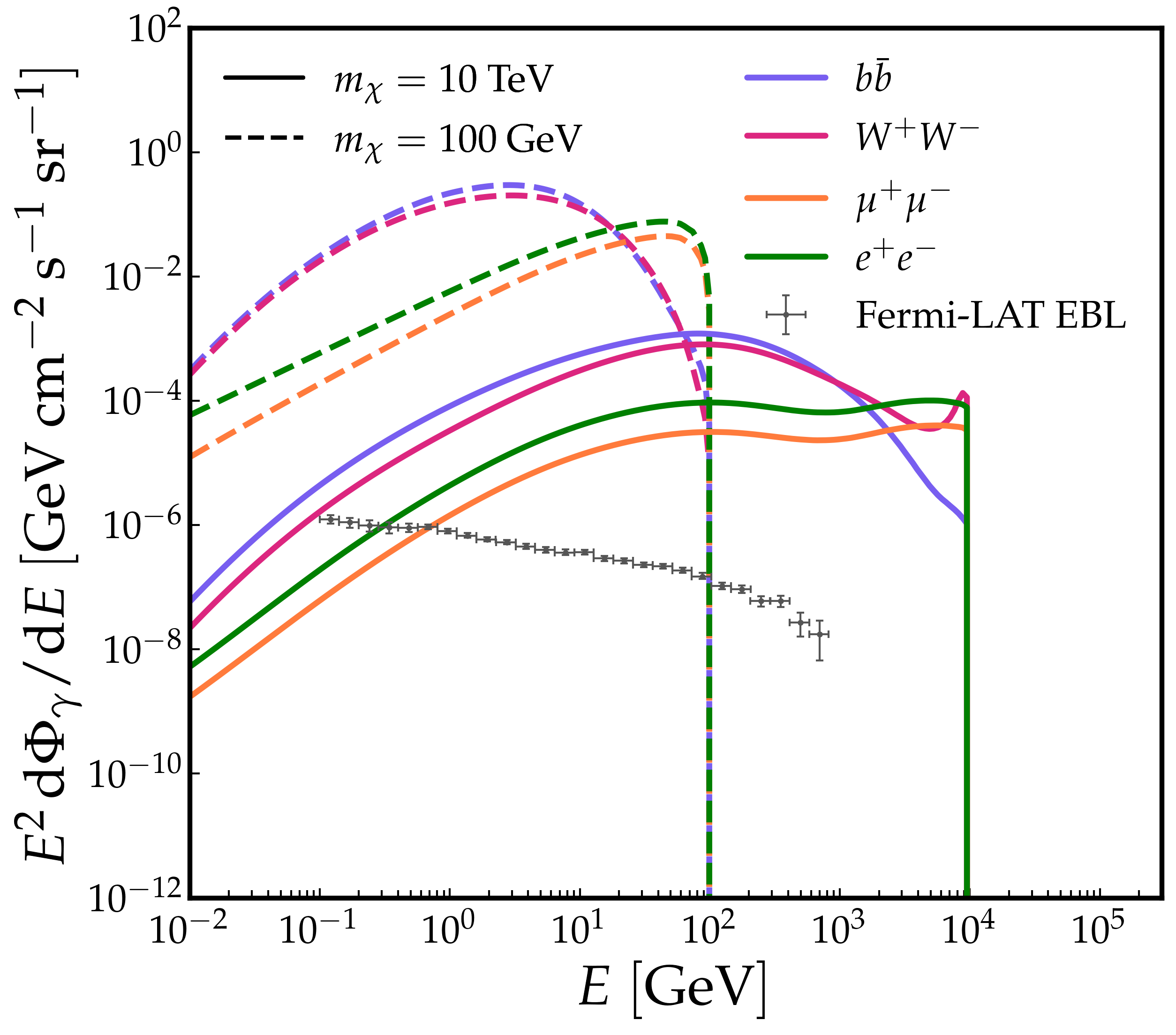}
    \end{subfigure}
    \hfill
    \begin{subfigure}[t]{0.495\textwidth}
        \centering
        \includegraphics[width=\linewidth]{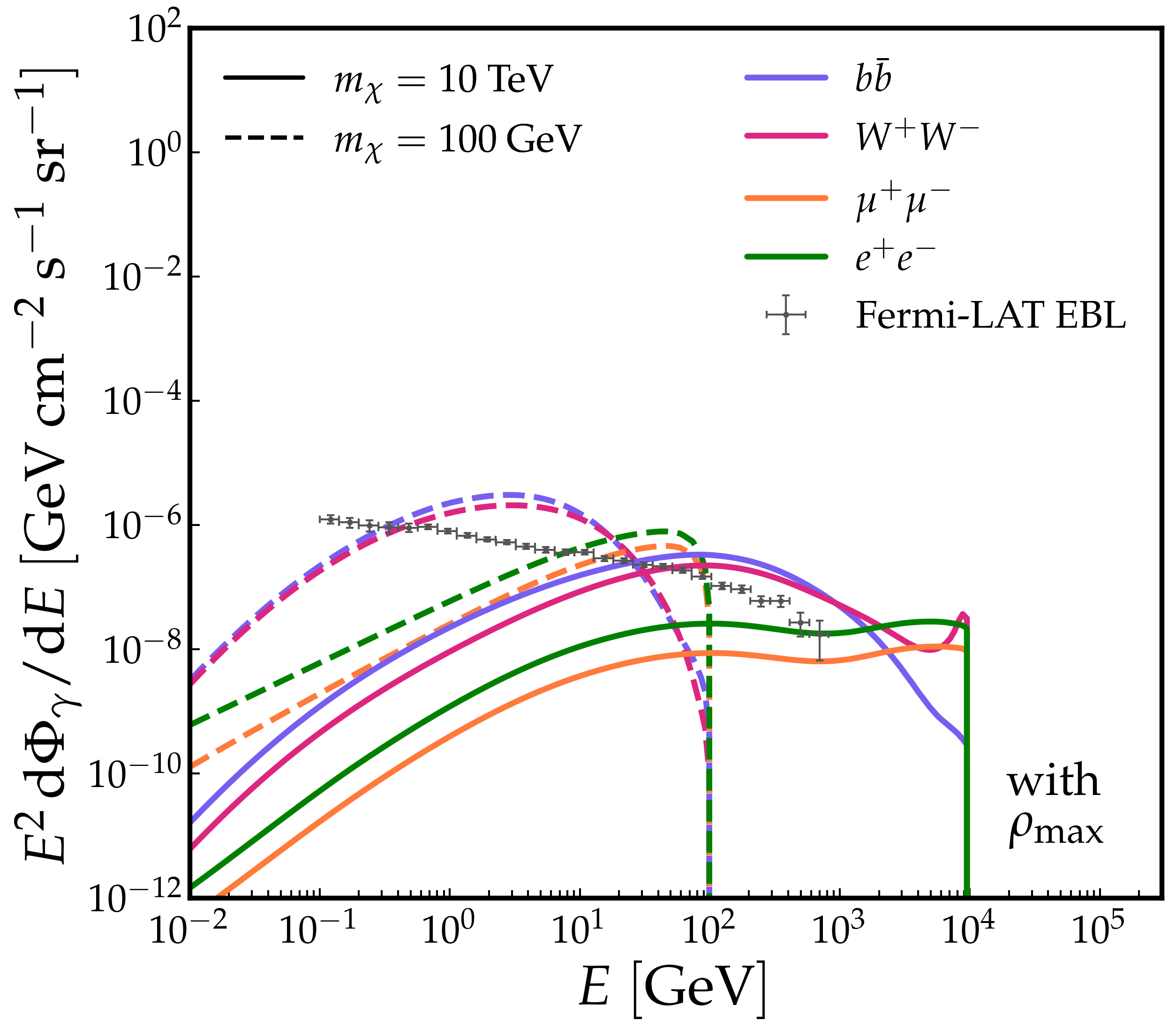}
    \end{subfigure}

\caption{
\textbf{Left:} Diffuse photon flux from BH DM spikes with constant density.
\textbf{Right:} Same, but including a plateau at $\rho_{\max}$ stemming from spike depletion due to annihilations.
Curves are shown for $m_\chi = 100$~GeV (dashed) and $m_\chi = 10$~TeV (solid) for the four annihilation channels considered. Black points denote the Fermi-LAT extragalactic $\gamma$-ray background used for comparison.
All fluxes assume the thermal relic cross-section $\langle\sigma v\rangle=3\times10^{-26}~{\rm cm^3~s^{-1}}$. 
}
    \label{fig:DS_flux_optdepth}
\end{figure}

We consider that the comoving BH population directly inherits the abundance and halo distribution of DSs at the collapse redshift $z_{\rm lim}=15$. Each DS present at that epoch leaves behind one BH of comparable mass. In our simplified picture we assume that the BHs do not subsequently significantly accrete, merge, or modify their host environment, and their spikes remain static over cosmic time as described in Sec.~\ref{sec:BH_profile}. Then, the total comoving luminosity density from the BH population can be estimated as
\begin{equation}\label{eq:BH_total}
    L_{\mathrm{BH}}^{\rm pop} = \int_{t(z_{\rm lim})}^{t(z)} dt'
    \int_{M_{h,\min}}^{M_{h,\max}} dM_h  
    L_{\mathrm{BH}}(M_h)  \frac{d f_{\rm surv}(z)}{dz} 
    \frac{d^2n_{\rm DS}}{dM_hdt'}\left(M_h,z(t-t')\right)  ,
\end{equation}
where $f_{\rm surv}(z)$ is defined in Eq.~\eqref{survival} and specifies the DS population at the time of collapse.

For the step function choice in Eq.~\eqref{survival}  one has
$d f_{\rm surv}/dz=\delta(z-z_{\rm lim})$ and thus Eq.~\eqref{eq:BH_total} appropriately selects the DS population at the moment of collapse. 
A more gradual transition survival function, as considered in Ref.~\cite{Schwemberger:2024rvu}, can be modeled as
\begin{equation} 
f_{\rm surv}^{\rm grad}(z) = \tfrac{1}{2}\left[1+\tanh  \left(\frac{z- z_0}{\Delta z}\right)\right].
\end{equation}
Its derivative is then
$
d f_{\rm surv}^{\rm grad}/dz
= 1/(2\Delta z) {\rm sech}^2  \left((z- z_0)/(\Delta z)\right),
$
which behaves as a smeared delta function centered at $z_0$. 
Thus, the structure of Eq.~\eqref{eq:BH_total} remains valid and the mapping between the DS and BH populations is analogous.
Since the BH masses and their spike profiles are taken to be time independent, the luminosity density $L_{\mathrm{BH}}^{\rm pop}$ does not evolve with redshift, although in a fully realistic treatment this will not be the case.

The diffuse photon flux observed today from BH spike annihilations is obtained by integrating the emission from the remnant population over redshift 
\begin{equation}\label{eq:BH_flux}
\frac{d\phi_\gamma^{BH}}{dE}(E) =
\int_{0}^{z_{\rm lim}} dz \left|\frac{dt}{dz}\right| 
\frac{(1+z)}{\langle E\rangle} 
L_{\mathrm{BH}}^{\rm pop} \frac{dN_\gamma}{dE}\bigg|_{E'=(1+z)E}
e^{-\tau(E,z)} ,
\end{equation}
where $dN\gamma/dE$ is the photon spectrum per annihilation, evaluated at the redshifted energy $E'=(1+z)E$, and the factors $\langle E\rangle$ and $f_\gamma$ are the channel dependent mean photon energies and emitted photon fractions (see Tab.~\ref{tab:fgamma}), respectively. The exponential factor accounts for attenuation on the EBL.

Fig.~\ref{fig:DS_flux_optdepth} shows the resulting diffuse flux in our two limiting prescriptions for the spike, a more conservative case in which the spike density is truncated by annihilation depletion at $\rho_{\max}$, and a case in which the spike remains static. For the benchmark thermal relic cross-section, both scenarios exceed the Fermi-LAT extragalactic background over broad energy ranges and thus result in new constraints. The spectral shapes follow the patterns discussed previously, see Fig.~\ref{fig:spectra} and Tab.~\ref{tab:fgamma}, while the impact of the optical depth is more moderate here than in the DS contracted halo overdensities because the emission originates at lower redshifts when $z \lesssim 15$. Consequently, strong absorption features are less pronounced  and the spectra extend to higher energies with limited redshifting. 

\section{Constraints on Dark Matter Annihilation}\label{sec:constraints}

\begin{figure}[t]
    \centering 
    \begin{subfigure}{0.495\textwidth}
        \centering
        \includegraphics[width=\linewidth]{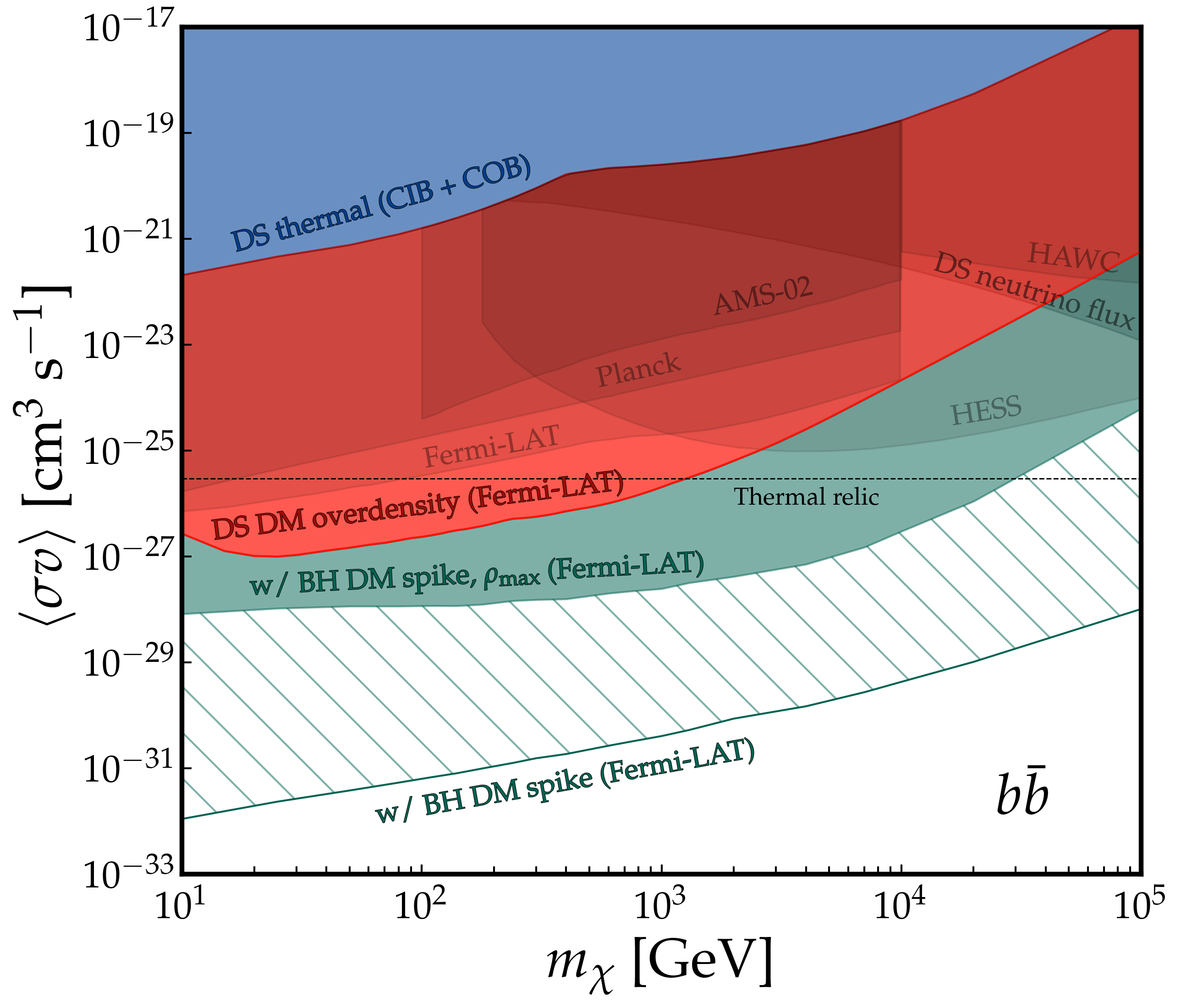}
    \end{subfigure}
    \begin{subfigure}{0.495\textwidth}
        \centering
        \includegraphics[width=\linewidth]{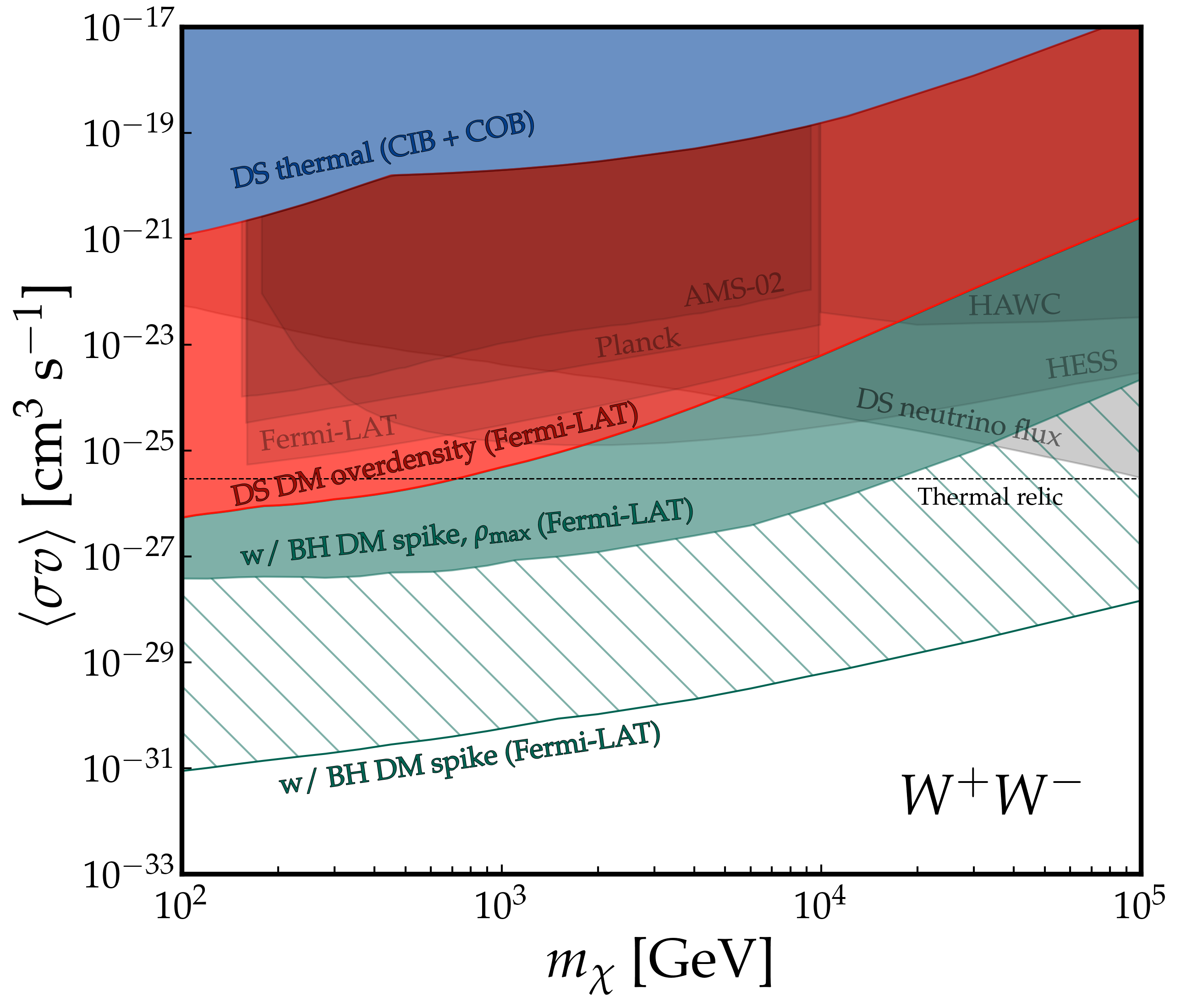}
    \end{subfigure} 
    \vspace{0.2cm}
    \begin{subfigure}{0.495\textwidth}
        \centering
        \includegraphics[width=\linewidth]{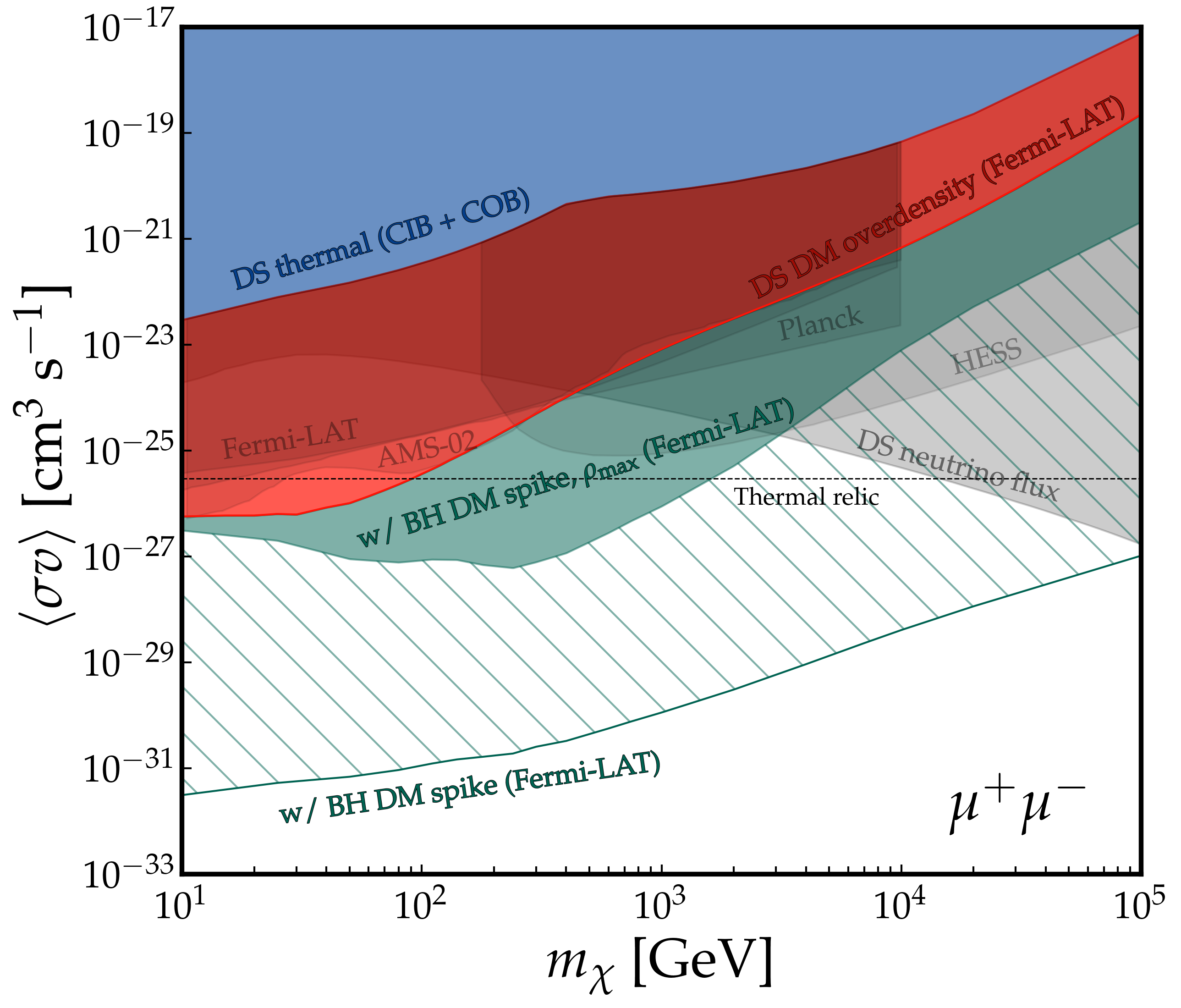}
    \end{subfigure}
    \begin{subfigure}{0.495\textwidth}
        \centering
        \includegraphics[width=\linewidth]{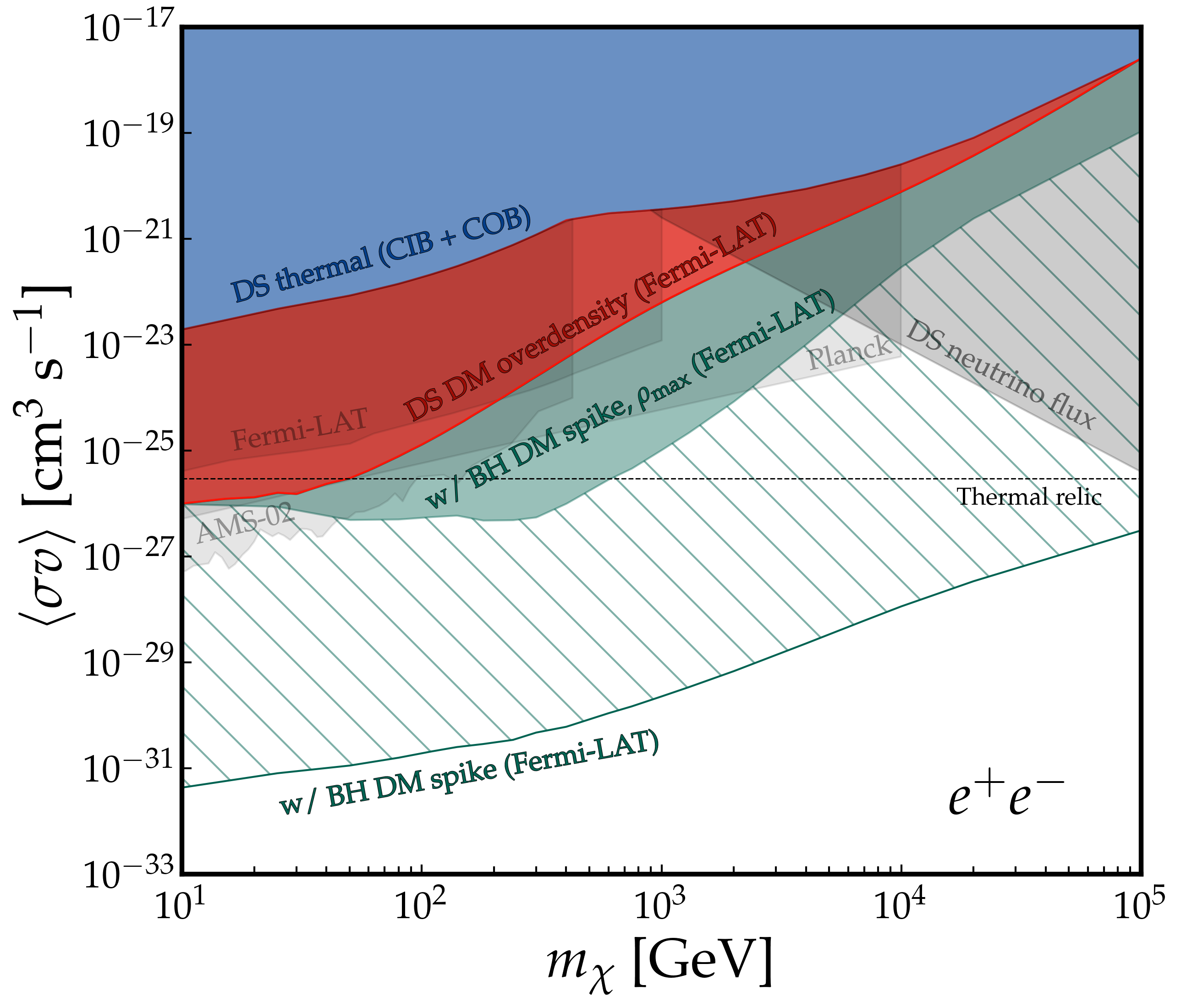}
    \end{subfigure}
 \caption{
Constraints on DM annihilation for the representative channels considered in this work $b\bar b$ (Top Left), $W^+W^-$ (Top Right), $\mu^+\mu^-$ (Bottom Left) and $e^+e^-$ (Bottom Right).
The shaded regions correspond to novel constraints from this work for the different emission components of our scenario including thermal emission from the DS surface constrained by CIB and COB data, annihilation in the contracted halo outside the DS  as well as the post-collapse BH spike contribution constrained by Fermi-LAT data.
The hashed band denotes constraints for the case of a static spike without depletion no $\rho_{\rm max}$ density cutoff, while the green shaded region shows the corresponding bounds when the DM annihilation depletion $\rho_{\rm max}$ is included.
DS diffuse neutrino background constraints from Ref.~\cite{Schwemberger:2024rvu}, together with new neutrino constraints for $e^+e^-$ channel from this work, are shown for comparison and highlight the complementarity between photon and neutrino probes of the same supermassive DS population.
The grey shaded regions indicate constraints from other indirect-detection searches including
HESS~\cite{HESS:2016mib},
HAWC~\cite{HAWC:2023owv},
Fermi-LAT~\cite{Fermi-LAT:2015att},
AMS-02~\cite{AMS:2014xys, AMS:2014bun} 
and CMB limits from Planck~\cite{Kawasaki:2021etm}.}
    \label{fig:constraints}
\end{figure}

Using the computed fluxes  we derive novel constraints on the DM microphysical parameters required to power a population of DSs capable of producing observable signatures. We adopt a conservative and simplified approach such that any DM model predicting a DS-induced flux that exceeds the measured diffuse astrophysical backgrounds by more than one standard deviation is considered excluded. We do not attempt to interpret or fit any potential excesses in the astrophysical diffuse photon backgrounds, which are typically subject to substantial modeling uncertainties. Instead, we use the observational data   to set  upper limits. For the parameter space of interest, we find that the most stringent bounds arise from the extragalactic $\gamma$-ray background measured by Fermi-LAT~\cite{Fermi-LAT:2015att}. Our constraints do not rely on detailed modeling of astrophysical background sources or on exploiting the full spectral information, both of which could strengthen our limits further.

In Fig.~\ref{fig:constraints} we summarize the resulting constraints.
We find that the enhanced DM densities required to power supermassive DSs already yield strong bounds from annihilations occurring outside the stellar surface and for the thermal relic cross-section DM masses below $\sim 1$~TeV for the $\chi\chi\to b\bar b$ and $W^+W^-$ channels, and below $\sim 100$~GeV for leptonic channels, are excluded.
At higher masses signals associated with DM annihilations in contracted DM halos during DS-phase are strongly attenuated by the optical depth. Then, the dominant constraints instead arise from annihilations in the DM spikes surrounding the BH remnants. These spike related limits, however, carry substantial theoretical uncertainties, consistent with the broader literature on DM spike evolution unrelated to DSs~\cite{Ahn:2007ty,Bertone:2005hw,Bertone:2005xv,Bertone:2005xz,Merritt:2003qk,Ullio:2001fb}.
We quantify this  uncertainty on the DM spike constraint by showcasing two scenarios, considering spikes with and without an annihilation-induced maximal density cut off $\rho_{\rm max}$. The cutoff is motivated by depletion from DM annihilation  and we adopt a conservative implementation by applying it from the moment of BH formation at $z_{\rm lim}=15$. These approximations are further discussed in Sec.~\ref{sec:BH_profile}.  
Finally, we note that the thermal surface emission of DSs is never competitive with these bounds and remains well below existing indirect-detection constraints.

Further, in Fig.~\ref{fig:constraints} we also compare our new electromagnetic diffuse flux constraints with those obtained for the same population of supermassive DSs from the diffuse neutrino background previously derived in Ref.~\cite{Schwemberger:2024rvu} by some of us. As discussed in App.~\ref{app:neutrinos}, the same DM annihilation channels can copiously produce neutrinos resulting in sizable neutrino fluxes  and unlike photons cosmological neutrinos do not suffer attenuation during propagation. For the $e^+e^-$ channel, which was not treated in Ref.~\cite{Schwemberger:2024rvu}, we derive the corresponding new neutrino limits here. This comparison highlights the multimessenger complementarity between photons and neutrinos, demonstrating that a broad range of DS scenarios capable of seeding SMBHs can be jointly probed by both channels with distinct detectors.

In App.~\ref{app:ST} we consider the effect of different halo formation models while in App.~\ref{app:collapse} and~\ref{app:profile} we consider the effect of different collapse redshifts $z_{\rm lim}$ and BH spike power-law indices respectively. These signify uncertainties on our results.

\section{Conclusions}
\label{sec:conclusions}

In this work we have derived new constraints on the particle physics properties of DM required to power a cosmological population of supermassive DSs capable of seeding SMBHs. Our study incorporates the full multimessenger emission history of DSs, from their initial formation through collapse. We include thermal surface emission, non-thermal annihilation signals from the adiabatically contracted halo surrounding the star  and the late-time contribution from the DM spike around the BH remnant.

Our analysis provides the first population integrated  diffuse multimessenger constraints on supermassive DSs powering SMBH formation, extending the reach of DS searches well beyond the redshift range directly accessible to instruments such as JWST. While some of the previous works have focused primarily on identifying individual supermassive DSs in high redshift surveys, our approach exploits the fact that a cosmological DS population produces an integrated electromagnetic background that remains observable even when individual objects are no longer visible. For DS masses $M_{\rm DS}\gtrsim 10^6 M_\odot$, where luminosities and lifetimes are naturally enhanced, we obtain strong constraints on DM microphysics across a wide class of DS SMBH seeding scenarios.

We find that the adiabatically contracted DM halos required to sustain supermassive DSs generate substantial diffuse $\gamma$-ray emission. For DM annihilation to hadronic or electroweak channels the external contracted halo region alone allows excluding thermal relic cross-sections for $m_\chi \lesssim {\cal O}(1~{\rm TeV})$, while purely leptonic channels are constrained below $\sim 100~$GeV. These bounds hold even without including the effects of post-collapse BH DM spike. At higher masses the DS-phase signal is suppressed by cosmological attenuation, and the dominant constraint arises from the BH spike. We cover the associated theoretical uncertainties by considering both static spikes without evolution and spikes truncated at the maximum density by DM annihilation depletion. Even in the more conservative depleted case  we find that constraints can be significant. In all scenarios that we consider  thermal surface emission remains subdominant relative to both annihilation components and to existing indirect DM detection bounds.

Our diffuse photon constraints are strongly complementary to those from the diffuse DS neutrino background. Neutrinos, unaffected by cosmological attenuation, can probe the highest DM masses, while photons provide the strongest sensitivity at lower masses. For the $e^+e^-$ channel, which had not been previously analyzed in this context, we have derived new neutrino limits that further highlight the multimessenger feasibility of our framework.

Our results indicate that DS-powered SMBH formation is subject to strong constraints from existing astrophysical extragalactic background observations over a range of particle DM scenarios. For DM particles with thermal-relic cross section  it is challenging for supermassive DSs surviving to $z\sim 15$ to exist in sufficient abundance to seed the bulk of the SMBH population if DM mass is $m_\chi \lesssim 10^2-10^3$~GeV, depending on the DM annihilation channel and spike evolution. Lighter, smaller or less abundant DSs are viable, and motivate further exploration of early Universe stellar formation in connection to DM. More broadly, our work establishes diffuse, multimessenger backgrounds as a powerful probe of novel stellar populations in the early Universe and their underlying particle physics connections.

\section*{Acknowledgment}
\addcontentsline{toc}{section}{Acknowledgments}

We thank Katherine Freese and Cosmin Ilie for discussions. This work was supported by World Premier International Research Center Initiative (WPI), MEXT, Japan. V.T. acknowledges support by the JSPS KAKENHI grant No. 23K13109. M.M. acknowledges the hospitality of the International Center for Quantum-field Measurement Systems for Studies of the Universe and Particles (QUP, WPI) during his visit, where this work was carried out. 

\appendix

\section{Neutrinos from dark matter annihilation}\label{app:neutrinos}

As discussed in Sec.~\ref{sec:polytrope}  not all of the DM annihilation energy is deposited within the stellar interior. A non-negligible energy fraction $f_\nu$ can escape as neutrinos. The value of $f_\nu$ depends on the DM mass and annihilation channel.

\begin{figure}[t]
    \centering
    \includegraphics[width=0.5\linewidth]{images/spectra_neutrinos.png}
    \caption{
    Neutrino spectra per DM annihilation for the channels $b\bar b$, $W^+W^-$, $\mu^+\mu^-$, and $e^+e^-$ considered in this work. Dashed curves correspond to a DM mass $m_\chi = 100~\mathrm{GeV}$, while solid curves correspond to $m_\chi = 10~\mathrm{TeV}$. The spectra are normalized such that $\int dE  (dN_\nu/dE)$ equals the number of neutrinos emitted per annihilation.}
    \label{fig:spectra_neutrinos}
\end{figure}

In Fig.~\ref{fig:spectra_neutrinos} we display the neutrino spectra for each channel computed using data of Ref.~\cite{Cirelli:2010xx}. The change in spectral shape and hardness with increasing $m_\chi$ determines the neutrino energy fraction $f_\nu$ entering Eq.~\eqref{eq:L_dm}. From these spectra, the neutrino energy fraction can be computed as 
\begin{equation}
f_\nu = \frac{1}{2m_\chi}\sum_{\alpha=e,\mu,\tau}\int dE   E   \frac{dN_{\nu_\alpha}}{dE},
\end{equation}
where $dN_{\nu_\alpha}/dE$ is the neutrino yield per annihilation into flavor~$\alpha$.
For reference, in Tab.~\ref{tab:f_nu} we show the resulting values of $f_\nu$ and the corresponding average neutrino energy for the channels considered in this work. We note that a fraction of the total annihilation energy goes into rest mass of final state particles (e.g.  baryons), thus generically one has $f_\nu + f_\gamma < 1$.

\begin{table}[t]
    \centering
\begin{tabular}{c|c|c|c|c}
\hline\hline
Annihilation&
\multicolumn{2}{c|}{$m_\chi = 100~\mathrm{GeV}$} &
\multicolumn{2}{c}{$m_\chi = 10~\mathrm{TeV}$} \\
 channel       & $f_\nu$ & $\langle E\rangle$ (GeV) 
        & $f_\nu$ & $\langle E\rangle$ (GeV) \\
    \hline
        $b\bar{b}$     
            & 0.45  & 0.944 
            & 0.45  & 4.19 \\
        $W^+W^-$       
            & 0.50  & 1.28  
            & 0.52  & 9.32 \\
        $\mu^+\mu^-$   
            & 0.62  & 31.0  
            & 0.62  & 206 \\
        $e^+e^-$ 
            & $8.4\times10^{-5}$   & 1.86   
            & 0.04   & 110 \\
    \hline\hline
    \end{tabular}
    \caption{
Neutrino energy fraction $f_\nu$ and average energy $\langle E\rangle$ for different DM annihilation channels, considering $m_{\chi} = 100$~GeV and 10 TeV.}
    \label{tab:f_nu}
\end{table}

\section{Impact of UV background modeling on flux attenuation}
\label{app:UVmodels}

\begin{figure}[t]
    \centering
    \includegraphics[width=0.5\textwidth]{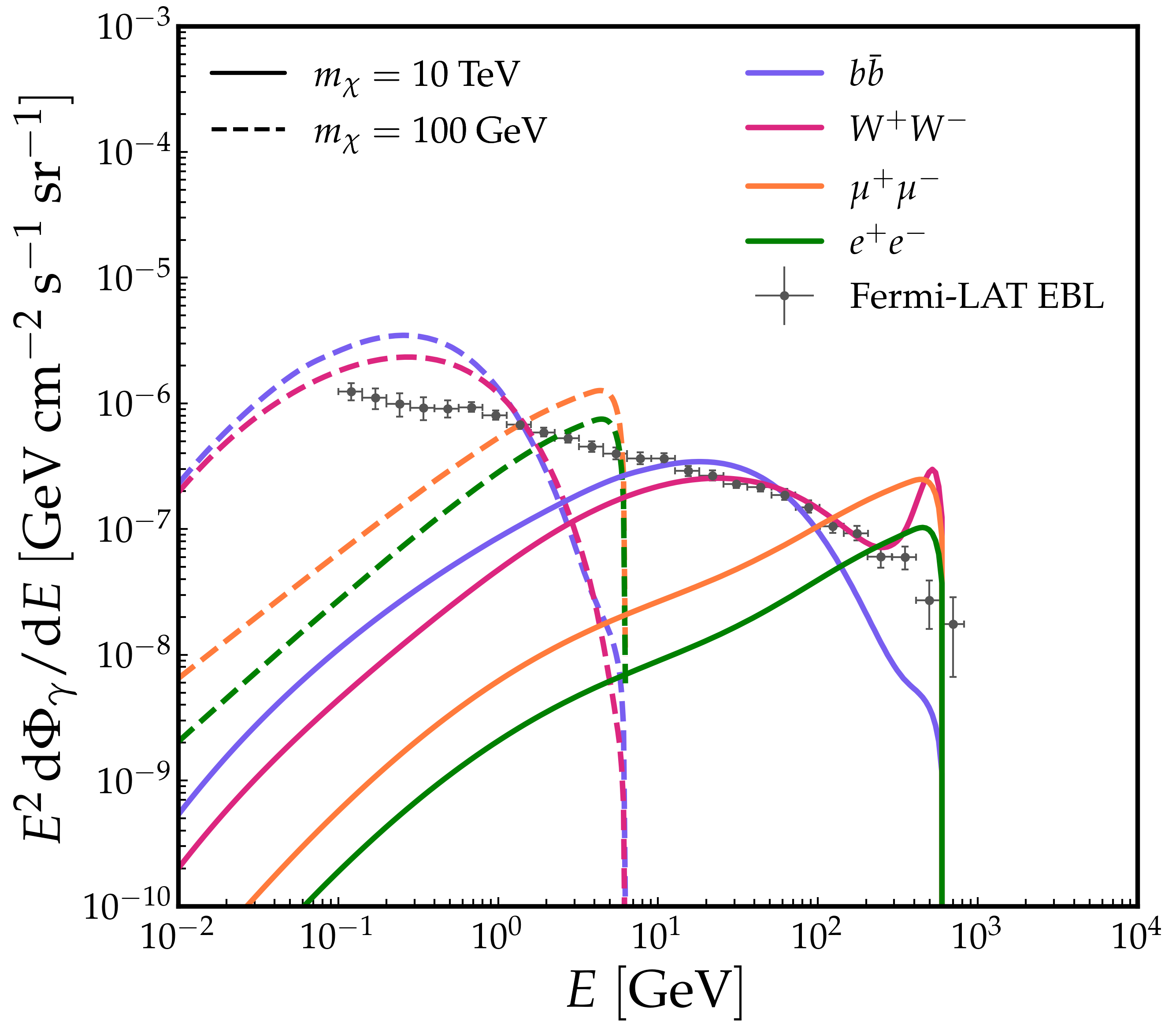}
    \caption{Diffuse photon flux from external to DS halo annihilations computed without cosmological attenuation.}
    \label{fig:Flux_noOPTICAL}
\end{figure}

\begin{figure}[t]
    \centering
    \begin{subfigure}{0.495\textwidth}
        \centering
        \includegraphics[width=\linewidth]{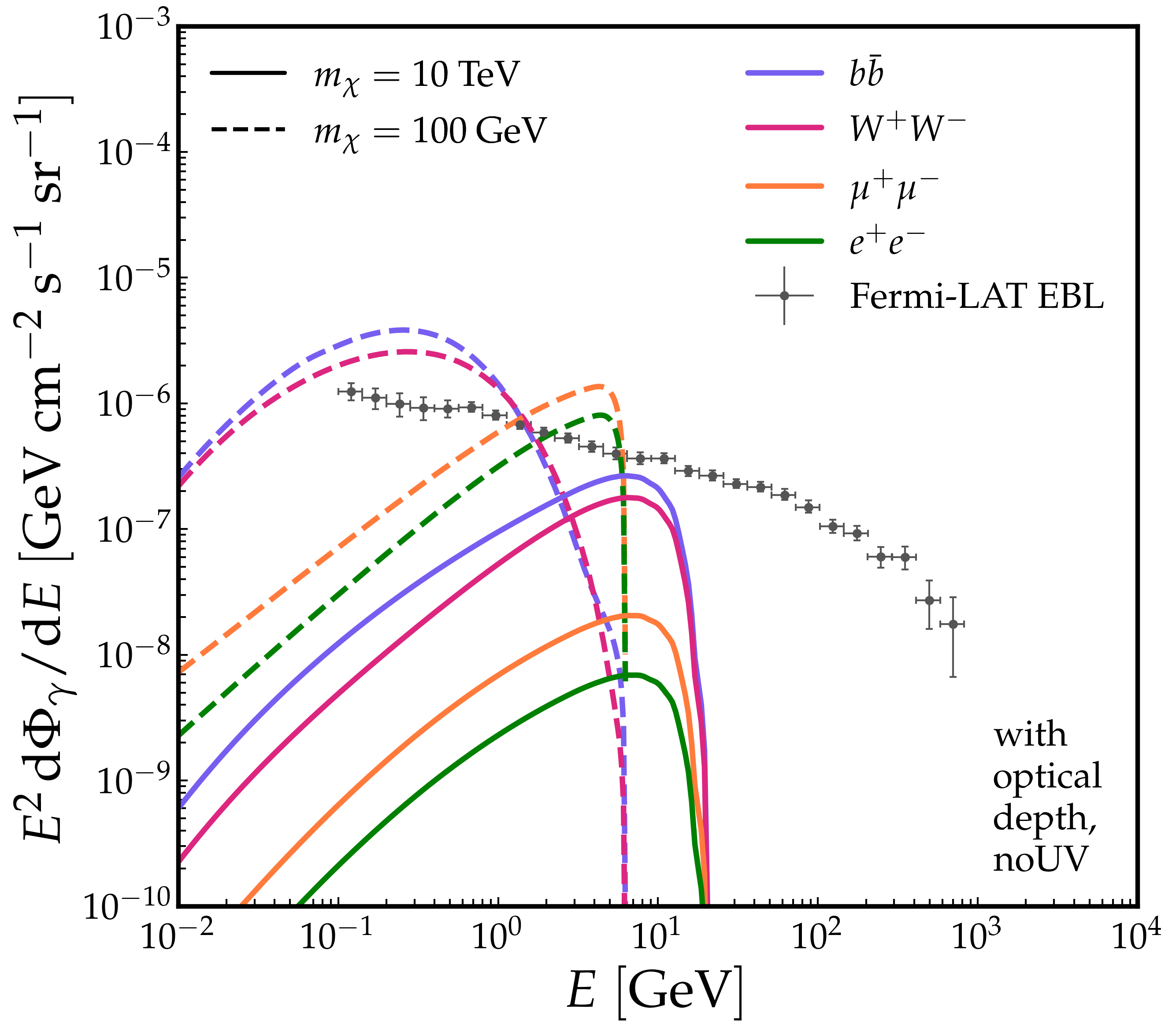}
    \end{subfigure}
    \hfill
    \begin{subfigure}{0.495\textwidth}
        \centering
        \includegraphics[width=\linewidth]{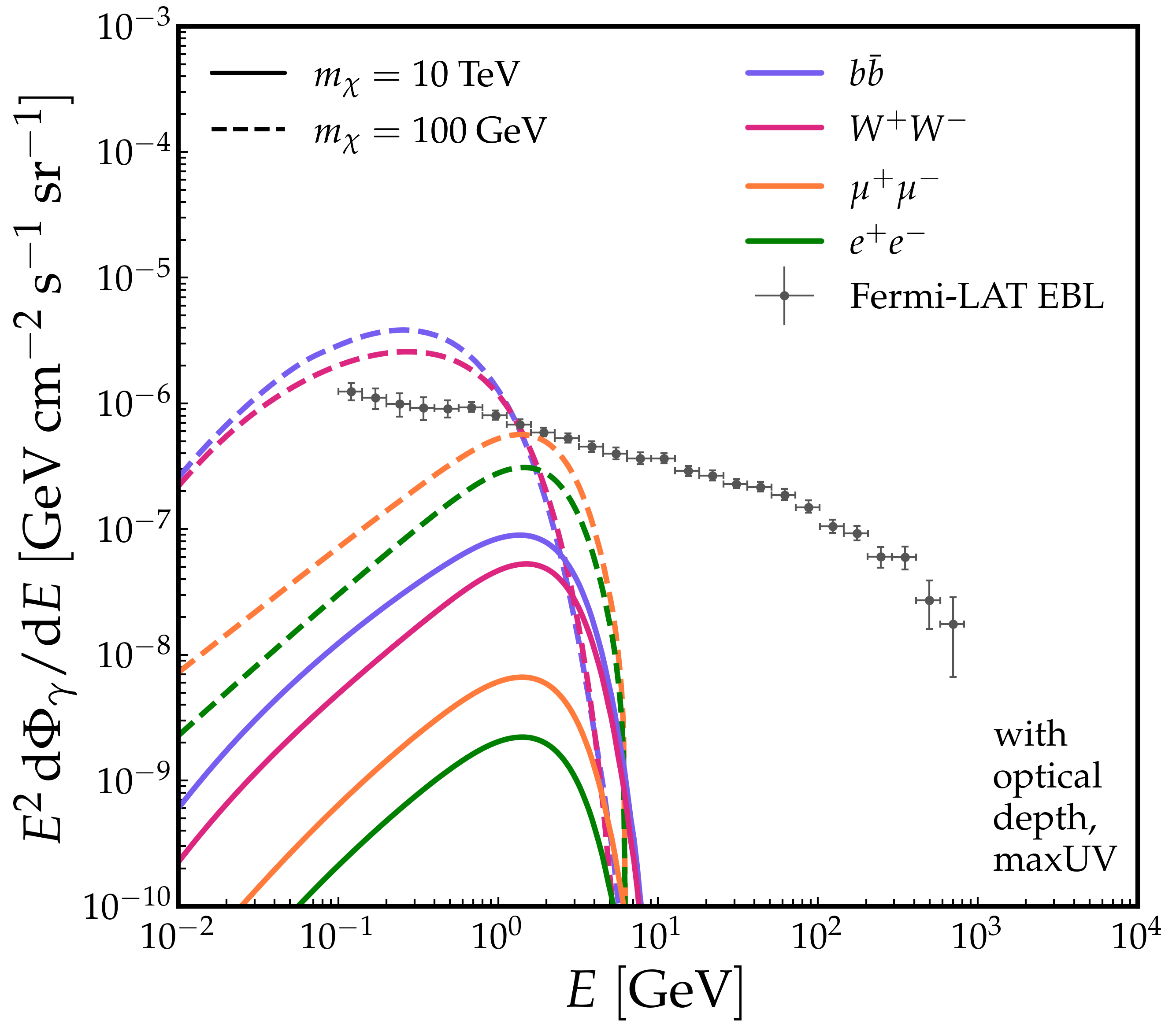}
    \end{subfigure}
     \caption{
Diffuse photon flux from external to DS halo annihilations for two extreme UV background prescriptions.  
    \textbf{Left:} noUV, corresponding to vanishing UV photon densities and minimal attenuation.  
    \textbf{Right:} maxUV, obtained by enhancing the minUV model by a factor of $\sim1.5$, resulting in the strongest UV-induced absorption.}
    \label{fig:Flux_UVcomparison}
\end{figure}

In Sec.~\ref{sec:DS_flux} we presented the diffuse photon flux from external DS annihilations using our reference optical depth prescription based on the ``minimal UV'' (minUV) model. Since the attenuation of high energy photons is sensitive to the ultraviolet component of the EBL, it is useful to examine how the predicted fluxes vary under different assumptions about the UV background. Here we compare several limiting scenarios, together with the idealized case in which cosmological attenuation is entirely neglected.

The UV portion of the EBL originates from the integrated emission of early stars and galaxies and remains one of the largest uncertainties in the optical depth calculation at the energies and redshifts relevant for this work. Following Ref.~\cite{Cirelli:2010xx}, we cover this uncertainty by considering three representative prescriptions. The noUV model sets the UV photon density to zero and corresponds to the least attenuated case. The minUV model adopts the minimal UV background compatible with current blazar attenuation measurements~\cite{MAGIC:2008sib}. Finally, the maxUV model enhances the minUV photon density by a factor of $\sim 1.5$  providing an upper bound on UV-induced absorption.

To illustrate the role of attenuation  in Fig.~\ref{fig:Flux_noOPTICAL} we show the diffuse flux from DM annihilation external to DSs obtained considering $e^{-\tau}=1$ for all energies and redshifts without any attenuation present. We compare this to the predictions obtained using the noUV and maxUV attenuation prescriptions in Fig.~\ref{fig:Flux_UVcomparison}.  
As expected, the noUV model yields the larger observable flux due to the absence of UV-induced absorption.  
Qualitatively similar effects occur  in the case of BH spikes as discussed in Sec.~\ref{sec:BH_flux}. In terms of constraints, using the noUV model strengthens the limits 
since the flux suffers minimal attenuation, while the maxUV scenario weakens them mildly due to the 
enhanced absorption since this is only a $\sim1.5$ increase over minUV. Overall, the minUV prescription provides a representative intermediate case.

\section{Emission comparison with uncontracted dark matter distributions}\label{app:NFW}

We compare, following Ref.~\cite{Schwemberger:2024rvu}, the photon flux generated by DM annihilations in the overdense regions surrounding DSs with the flux expected from standard halos following an NFW profile. This allows to quantify how dramatically adiabatic contraction around DSs boosts the annihilation signal relative to ordinary indirect DM detection scenarios.
Since the annihilation rate scales as $\rho_{\rm DM}^2$, we characterize the enhancement produced in the DS environment by the ratio
\begin{equation}
    f_{\rm DS}
    = \frac{\int_{R_s}^{r_h} \rho_{\rm \chi}^2(r)  r^2 \mathrm{d}r}
           {\int_0^{r_h} \rho_{\rm NFW}^2(r)  r^2 \mathrm{d}r} ,
\end{equation}
where $\rho_\chi$ is the contracted DM profile surrounding the DS, $R_s$ is the stellar photospheric radius, $r_h$ is the halo radius  and $\rho_{\rm NFW}$ is the NFW profile defined in Sec.~\ref{sec:BH_profile}.

To compute the reference NFW halo emission flux  we employ the same comoving emissivity formalism as in Sec.~\ref{sec:DS_flux}, except that the single halo luminosity is evaluated using $\rho_{\rm NFW}$ instead of the contracted profile and the full halo population is included. The corresponding comoving luminosity density is
\begin{equation}\label{eq:NFW_pop_lum}
    L^{\rm pop}_{\rm halo}(z)
    = \int_{M_{\min}}^{M_{\max}} dM_h \;
    L_{\rm halo}(M_h,z) 
    \frac{dn_h}{dM_h}(M_h,z)  ,
\end{equation}
where $dn_h/dM_h$ is the halo formation rate from Eq.~\eqref{eq:haloformationrate}.  
In contrast to the DS case where emission is restricted to the finite DS lifetime with $z_{\rm form}=25 \ge z\ge z_{\rm lim}=15$  the NFW contribution accumulates over the full considered cosmic history from $z_{\rm form}=25$ down to $z\simeq 0$.

\begin{figure}[t]
    \centering
    \includegraphics[width=0.5\linewidth]{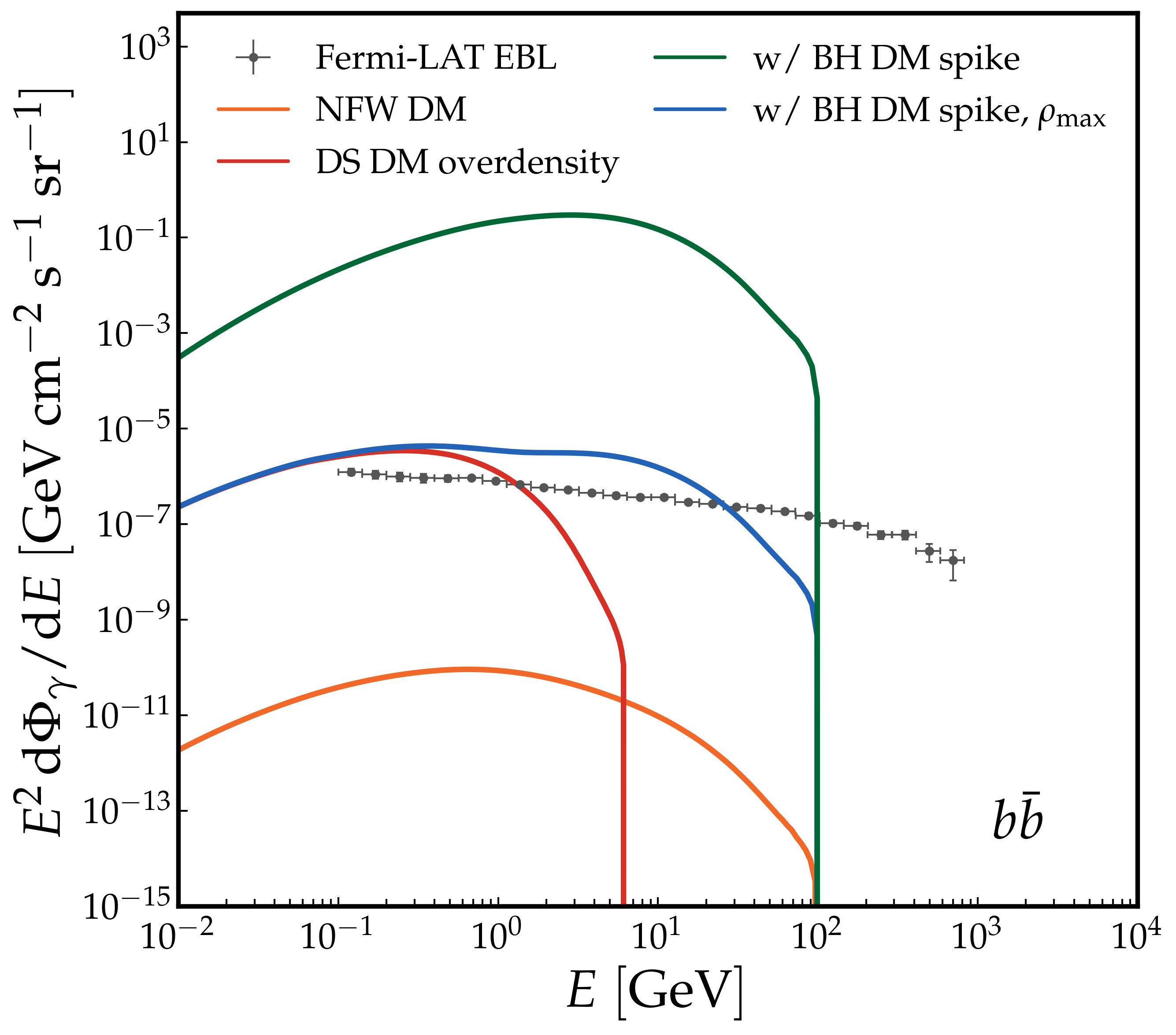}
\caption{
Photon flux from the components considered in this work.  
Orange curve depicts NFW halo contribution, integrated from $z_{\rm form}=25$ to $z\simeq 0$.  
Red curve depicts emission contribution from DM overdensities surrounding DS during their active phase $z_{\rm form}=25 \ge z\ge z_{\rm lim}=15$. For reference, we also display the additional photon flux contributions from BH-induced DM spikes forming
after DS collapse, with green denoting static spike without depletion and blue denoting spike with density limited to $\rho_{\rm max}$ by annihilation depletion. 
}
    \label{fig:indirect_det}
\end{figure}

In Fig.~\ref{fig:indirect_det} we compare  the photon flux from   NFW halos (orange) with the flux from DS-induced contracted halos (red). Their ratio directly illustrates the enhancement encoded in $f_{\rm DS}$. For reference, we also display the additional to contracted halo emission flux from BH-induced DM spikes formed after DS collapse, including both the static spike (green) and the annihilation-saturated spike with a $\rho_{\rm max}$ cutoff (blue). We note that the BH spike components are shown only for comparison and they do not enter the definition of $f_{\rm DS}$, which characterizes the enhancement from NFW to DS DM halo overdensities during the DS phase.
The benchmark spectrum shown corresponds to a $b\bar b$ annihilation channel with $m_\chi=100$~GeV, though the qualitative behavior is similar for other channels.

As we find, the DS-induced DM overdensity flux exceeds the NFW reference contributions by nearly $\sim10^{5}$, reflecting the large contraction  enhancement.
The magnitude of this enhancement is consistent with expectations from earlier studies of DS-induced annihilation signals. Early work on $\gamma$-ray backgrounds from $10$-$100~M_\odot$ DSs forming in $10^5$-$10^6~M_\odot$ halos~\cite{Yuan:2011yb}, based on the profiles of Ref.~\cite{Freese:2008hb}, found enhancement factors of only $f_{\rm DS}\sim10^3$, corresponding to early evolutionary phases with baryonic densities of order $\sim10^{13}~\mathrm{cm^{-3}}$. 
For the supermassive DSs considered in our work, the baryonic
contraction is far more extreme. As shown in Ref.~\citep{Rindler-Daller:2014uja},
DSs can reach central densities approaching
$\sim 10^{20}~\mathrm{cm^{-3}}$. Even an increase from
$10^{13}~\mathrm{cm^{-3}}$ to $10^{16}~\mathrm{cm^{-3}}$, holding the
halo mass fixed and taking the DS mass to be as small as
$M_{\rm DS}\sim 10^{-4} M_h$, enables increasing the annihilation
enhancement to $f_{\rm DS}\sim 10^6$. As the DS grows to
$\sim 1\%$ of the halo mass, the enhancement can rise to
$f_{\rm DS}\sim 10^8$. 

In our analysis we adopt a supermassive DS occupation fraction of $f_{\rm SMDS}=10^{-2}$, such that there is approximately one supermassive DS per hundred halos. Therefore, the effective enhancement entering Fig.~\ref{fig:indirect_det} is reduced from the intrinsic $f_{\rm DS}\sim10^8$ to $f_{\rm DS}\sim f_{\rm SMDS}\times10^8\simeq10^6$.  
The slightly smaller numerical enhancement in our calculations arises because the NFW contributions continue  to accumulate emission from $z<15$, whereas DS host DM halos contribute only during the DS epoch. The additional low redshift photon production from NFW halos therefore further reduces the relative contrast.

\section{Halo formation models}\label{app:ST}

The predicted DS and BH induced diffuse emissions depend    on the abundance and redshift distribution of the host halos in which supermassive DSs form. Throughout the main text we adopted the conventional Press-Schechter   formalism. Here we compare these results with the Sheth-Tormen  model~\cite{Sheth:1999mn}, which predicts a higher abundance of collapsed structures at early times.
This difference has a  physical impact on the DS diffuse flux.

\begin{figure}[t]
    \centering
    \begin{subfigure}[t]{0.495\textwidth}
        \centering
        \includegraphics[width=\linewidth]{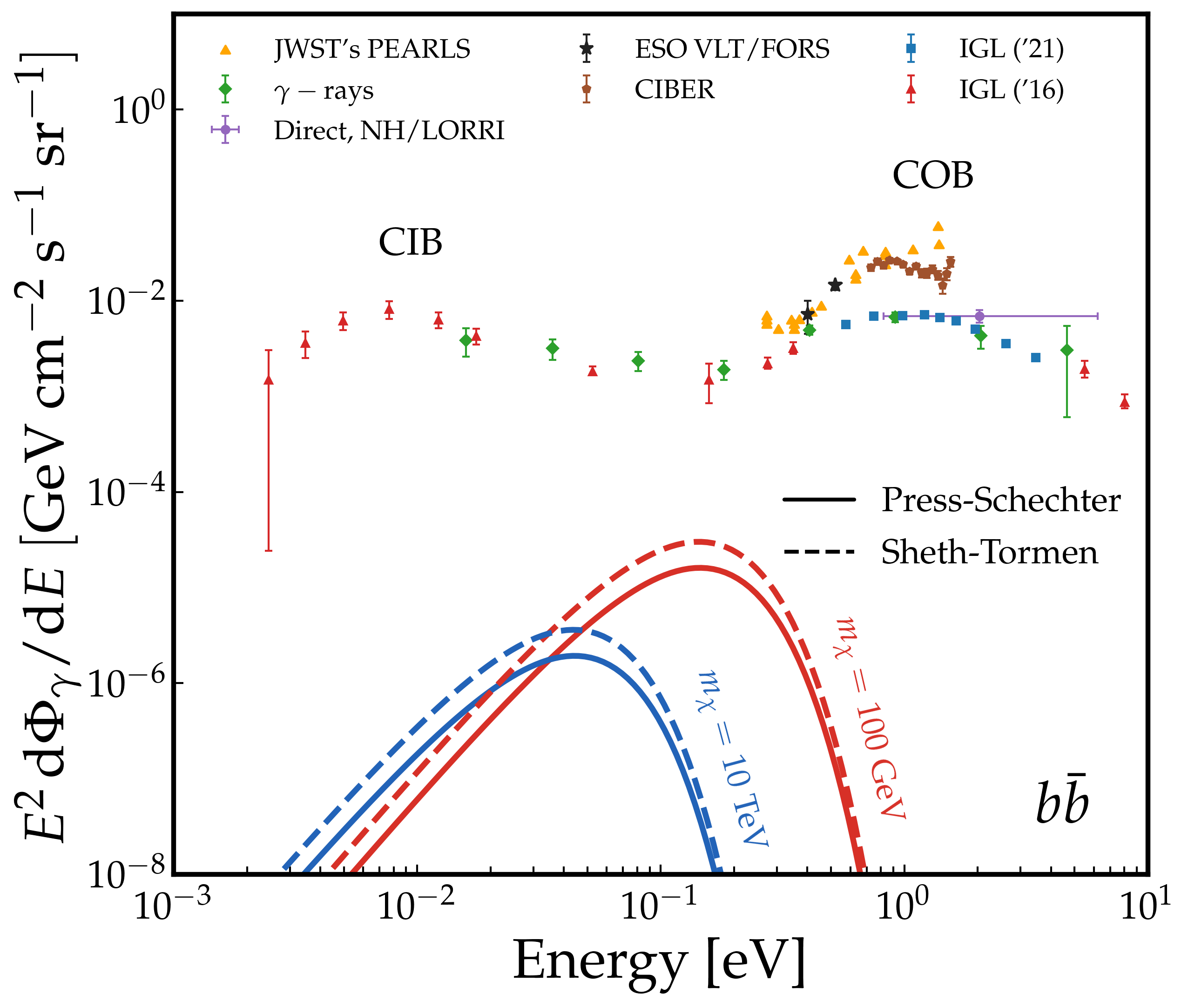}
    \end{subfigure}
    \hfill
    \begin{subfigure}[t]{0.495\textwidth}
        \centering
        \includegraphics[width=\linewidth]{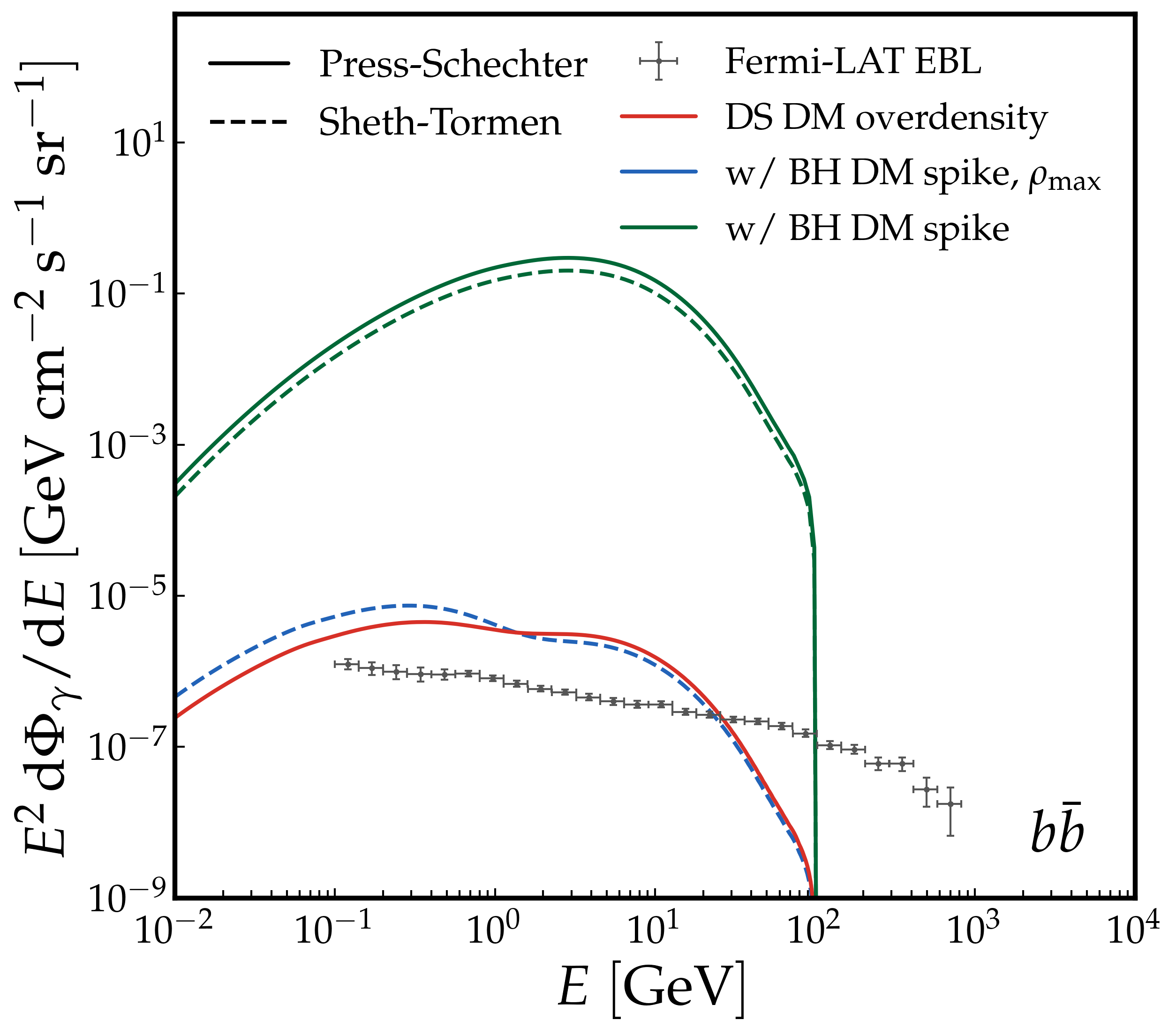}
    \end{subfigure}
\caption{
Comparison of the diffuse DS emission fluxes for the 
    $b\bar{b}$ annihilation channel considering
the Press-Schechter (PS, solid) and Sheth-Tormen (ST, dashed) 
halo formation models.  
\textbf{Left:} Thermal emission from the DS population.  
\textbf{Right:} Diffuse $\gamma$-ray flux from the DS external annihilation 
component (red) combined with contributions from BH spikes (blue and green). DM mass is considered to be $m_{\chi} = 100$ GeV.
}
    \label{fig:ST_comp}
\end{figure}

In the Sheth-Tormen model halos are both  more abundant  and form  earlier  at $z\gtrsim 20$ relative to the Press-Schechter prediction.  
Earlier formation enhances the number of DSs active at a given epoch and extends the duration of the accretion phase, thereby increasing the total DS luminosity.  
For our benchmark parameter choices, the net effect is an overall enhancement of the DS diffuse flux by a factor of few when adopting the Sheth-Tormen mass function.

The BH-spike contribution, by contrast, is only weakly sensitive to the choice of halo formation model.  
Because the BH-induced emission is accumulated from $z_{\rm lim}=15$ to present day, most of the signal originates at low redshift, where Press-Schechter and Sheth-Tormen predict nearly identical halo abundances.  
Moreover, since DSs generically reach their limiting mass well before collapse, the resulting BH masses are expected to be similar in both scenarios.  
A small difference stems from the fact that below $z_{\rm lim}$ the Sheth-Tormen mass function predicts a slightly reduced comoving abundance of halos in the narrow mass range relevant for DS collapse, leading to a mild suppression of the BH-spike flux relative to Press-Schechter.

\begin{figure}[t]
    \centering

\includegraphics[width=0.5\linewidth]{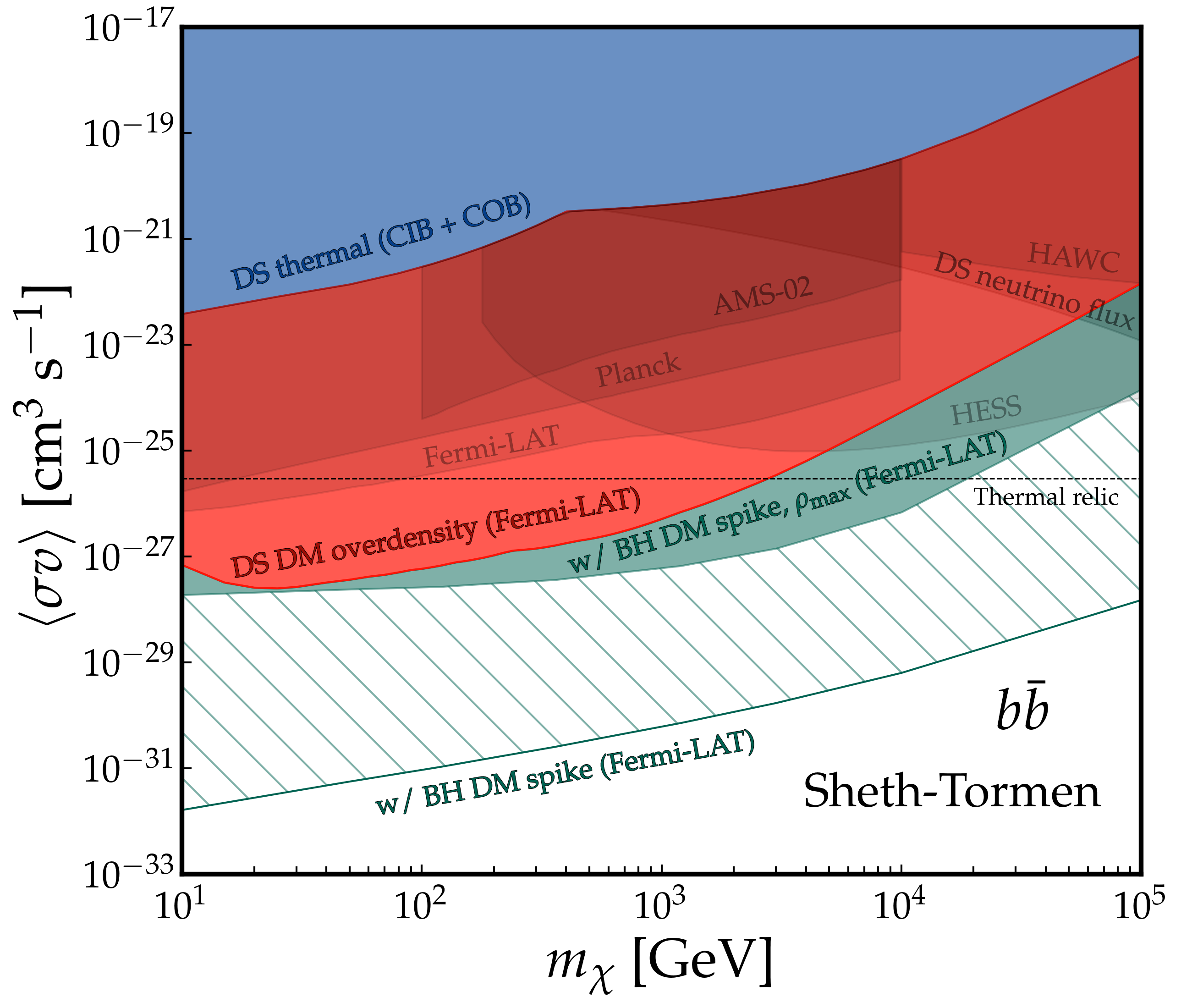}
    \caption{
Constraints on the DM annihilation cross-section for the $b\bar{b}$ channel 
    derived using the Sheth-Tormen mass function. 
    }
\label{fig:z_comp_constraints}
\end{figure}

In Fig.~\ref{fig:ST_comp} we illustrate these effects.  
The left panel compares the DS thermal emission in the Press-Schechter and Sheth-Tormen cases, while the right panel shows the corresponding diffuse $\gamma$-ray fluxes from DS external annihilations and BH spikes for the benchmark $b\bar{b}$ channel and $m_\chi = 100~\mathrm{GeV}$. 
 The resulting constraints considering Sheth-Tormen are shown in Fig.~\ref{fig:z_comp_constraints}, which can be compared directly with the Press-Schechter results in Fig.~\ref{fig:constraints}.  
For all other annihilation channels considered the relative behavior between the two models is qualitatively similar, the DS-induced DM halo contribution mildly increases in Sheth-Tormen, while the BH-spike constraints remain nearly unchanged except for a small suppression reflecting the slightly reduced halo abundance at $z_{\rm lim}$.

\section{Collapse of dark stars at different redshifts} \label{app:collapse}

In the main text we adopted a reference DS collapse redshift $z_{\rm lim}=15$. In our framework the rapid collapse is implemented through the survival factor $f_{\rm surv}(z)$ as in Eq.~\ref{survival}.
Here we consider the impact of delaying collapse to $z_{\rm lim}=10$.  
This choice requires a modification of our population model since below $z=15$ halo abundance growth is dominated by mergers, so keeping a fixed halo occupation fraction $f_{\rm SMDS}$ could result in far too many BH remnants by $z_{\rm lim}=10$.  
To avoid potentially artificially overproducing SMBH seeds, we assume a decoupled DS population from the evolving halo mass function and simply fix the DS  and resulting BH  abundance to factor $(f_{\rm SMDS} n_h)$ evaluated at $z_{\rm lim}=15$.  
While this simplified treatment neglects potential DS and BH mergers and implicitly assumes DSs survive halo mergers this allows a consistent comparison of distinct DS collapse redshifts. 

\begin{figure}[t]
    \centering
    \begin{subfigure}[t]{0.495\textwidth}
        \centering
        \includegraphics[width=\linewidth]{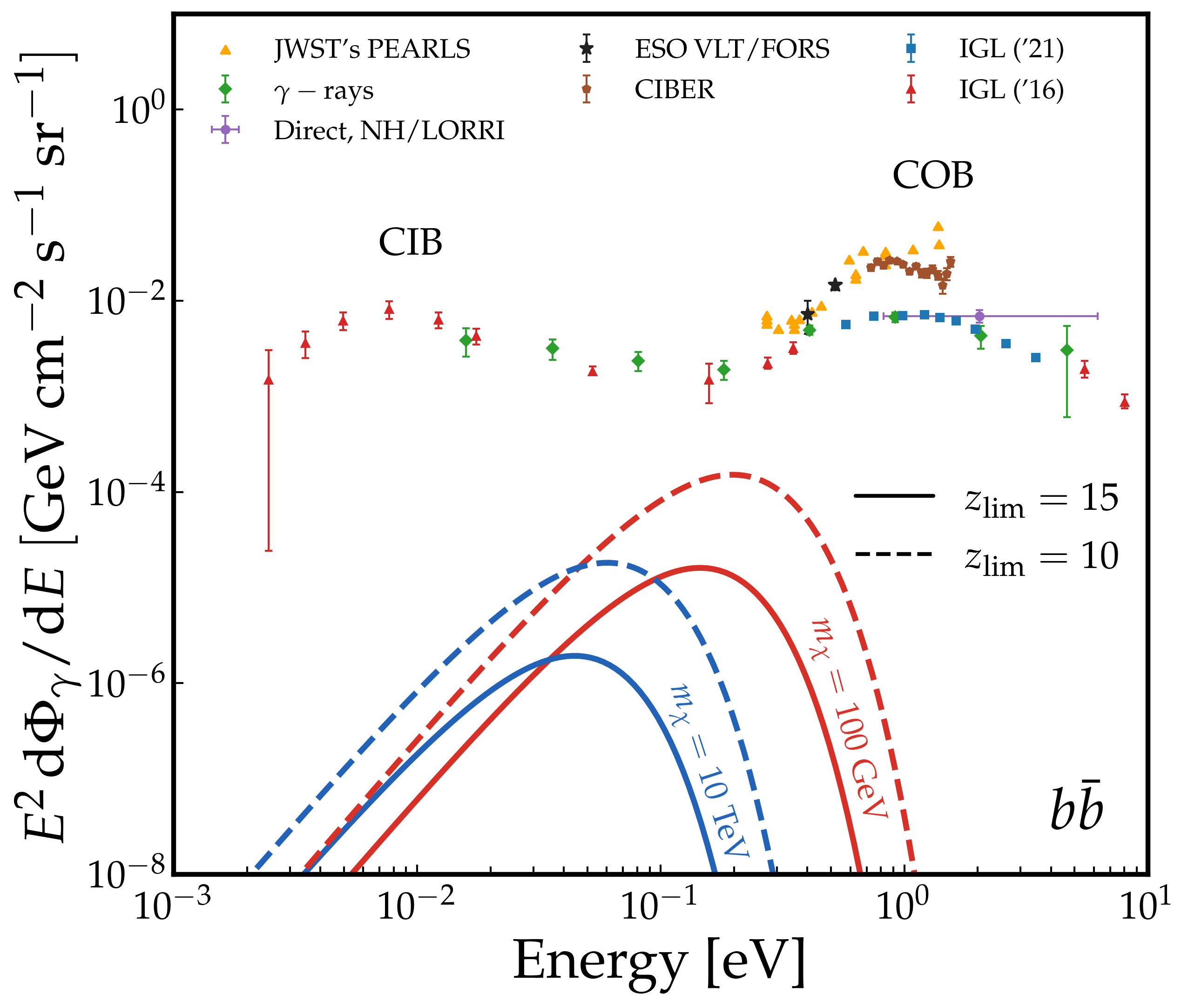}
    \end{subfigure}
    \hfill
    \begin{subfigure}[t]{0.495\textwidth}
        \centering
        \includegraphics[width=\linewidth]{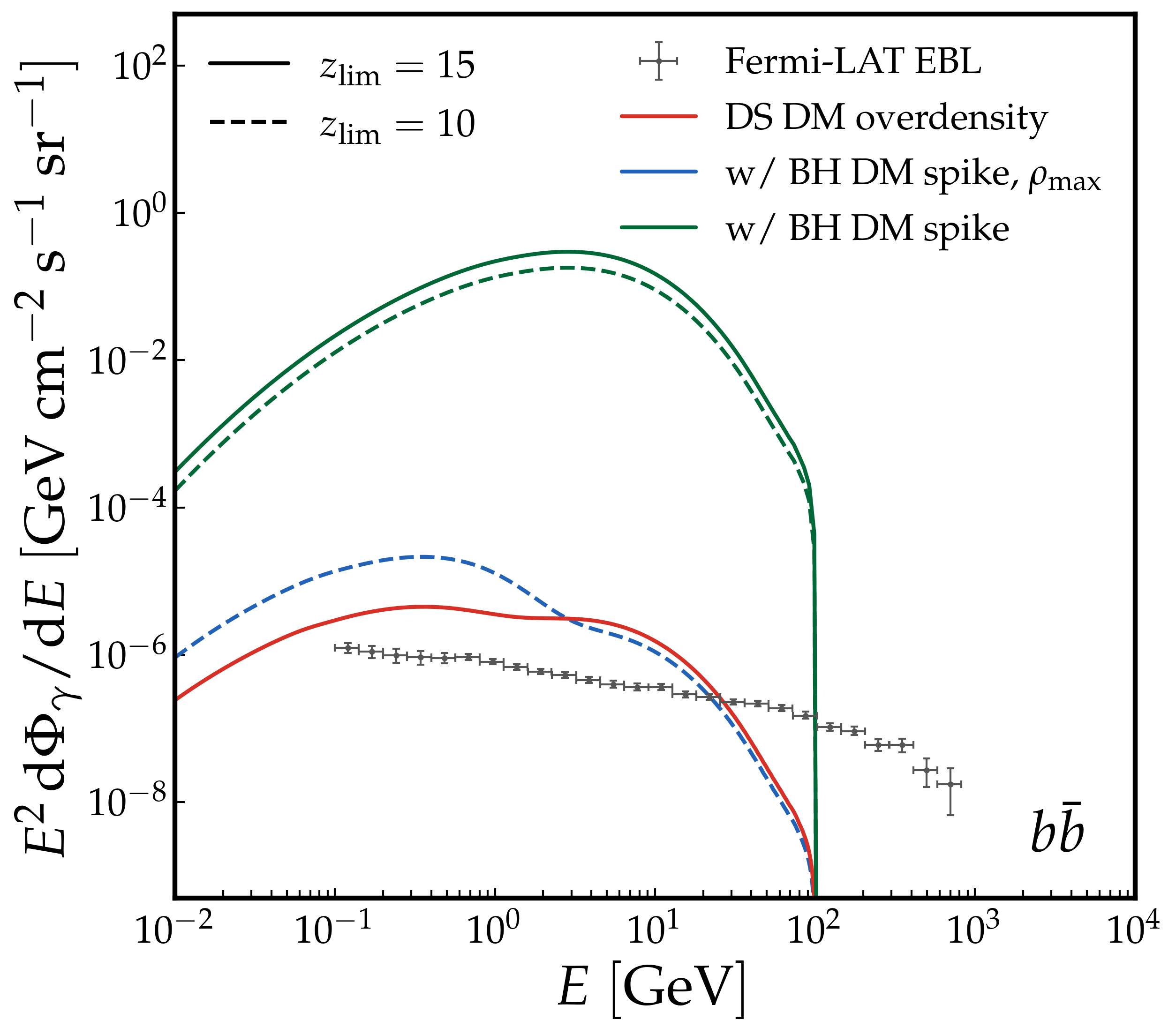}
    \end{subfigure}
    \caption{
    Comparison of the diffuse DS emission fluxes for the 
    $b\bar{b}$ annihilation channel obtained for different DS collapse 
    redshifts $z_{\rm lim}$. 
    \textbf{Left:} Thermal flux from the DS population for the 
    fiducial choice $z_{\rm lim}=15$ (solid) and the extended 
    lifetime case $z_{\rm lim}=10$ (dashed). 
    \textbf{Right:} Diffuse $\gamma$-ray flux from DS external 
    annihilation (purple) and BH spikes (green) for 
    $z_{\rm lim}=15$ (solid) and $z_{\rm lim}=10$ (dashed). DM mass is considered to be
    $m_\chi = 100~\mathrm{GeV}$. 
    }
    \label{fig:z_comp}
\end{figure}

\begin{figure}[t]
    \centering
    \includegraphics[width=0.5\linewidth]{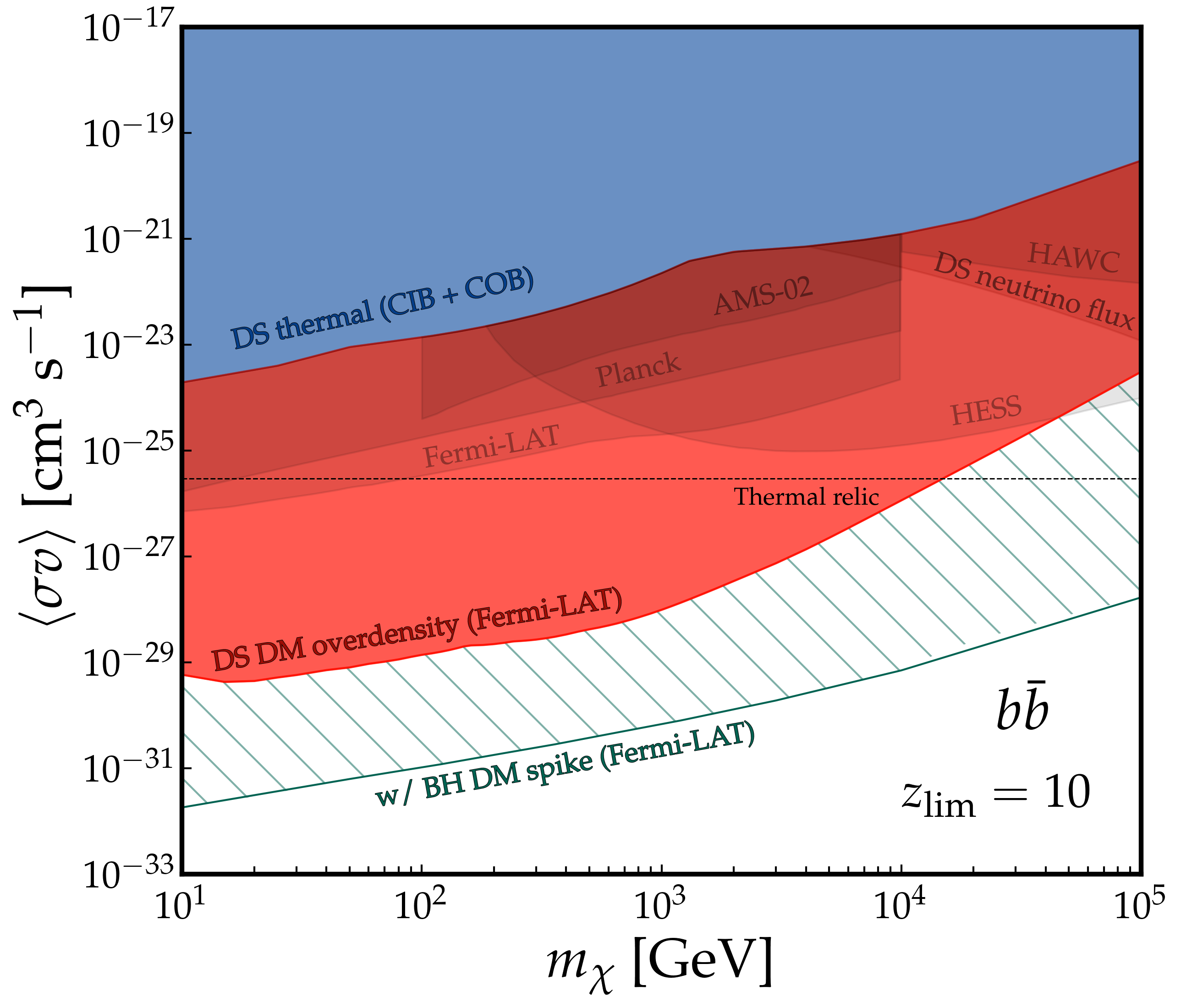}
    \hfill
    \caption{
Constraints on the DM annihilation cross-section for the $b\bar{b}$ channel assuming 
    $z_{\rm lim}=10$, analogous to   Fig.~\ref{fig:constraints}
    in the main text but with $z_{\rm lim}=15$.
    }
\label{fig:z_comp_constraints2}
\end{figure}

The resulting diffuse fluxes are shown in Fig.~\ref{fig:z_comp}.  
Lowering $z_{\rm lim}$ increases both the DS thermal and external DM halo annihilation contributions, since DSs radiate over a longer interval.  
Conversely, the BH-spike contribution is slightly reduced  as the BH phase begins later and therefore contributes over a shorter redshift range, although the effect is small because most of the BH-induced emission arises at $z\lesssim 5$.
In Fig.~\ref{fig:z_comp_constraints2} we show the impact on corresponding constraints.  
As expected, the DS external DM overdensity contribution becomes more restrictive, while the BH-spike limits are weakened due to the shortened epoch with BHs.  
We show only display the case of static BH-spike case, as the case limited by $\rho_{\max}$ never exceeds the DS overdensity contribution for $z_{\rm lim}=10$.  
We find that other DM annihilation channels exhibit similar qualitative behavior.

These results signify that that our constraints depend  on 
the assumed DS collapse redshift $z_{\rm lim}$.  
Lowering $z_{\rm lim}$ enhances DS-related emission and strengthens the resulting limits, while a higher $z_{\rm lim}$ would suppress DS-related contributions and correspondingly weakens the constraints.

\section{Dark matter spike profile}\label{app:profile}

The luminosity of BH-induced DM spikes carries sizable theoretical uncertainties, primarily because the inner density profile is sensitive to a number of poorly known dynamical processes.  
To quantify the impact of the spike shape itself, we vary the power law index in the parametrization $ 
    \rho_{\rm sp}(r) \propto r^{-\gamma_{\rm sp}}$  
considering $\gamma_{\rm sp}=1.5$ and $\gamma_{\rm sp}=9/4$ in addition to the reference $\gamma_{\rm sp}=7/3$ adopted in the main text.
For each spike profile we compute the corresponding luminosity with and without the annihilation depletion restriction on maximum density $\rho_{\max}$.  
We use the same benchmark configuration as in the main text considering annihilation into $b\bar b$ with $m_\chi = 100$ GeV.  

\begin{figure}[t]
    \centering
    \begin{subfigure}[t]{0.495\textwidth}
        \centering
         \includegraphics[width=\linewidth]{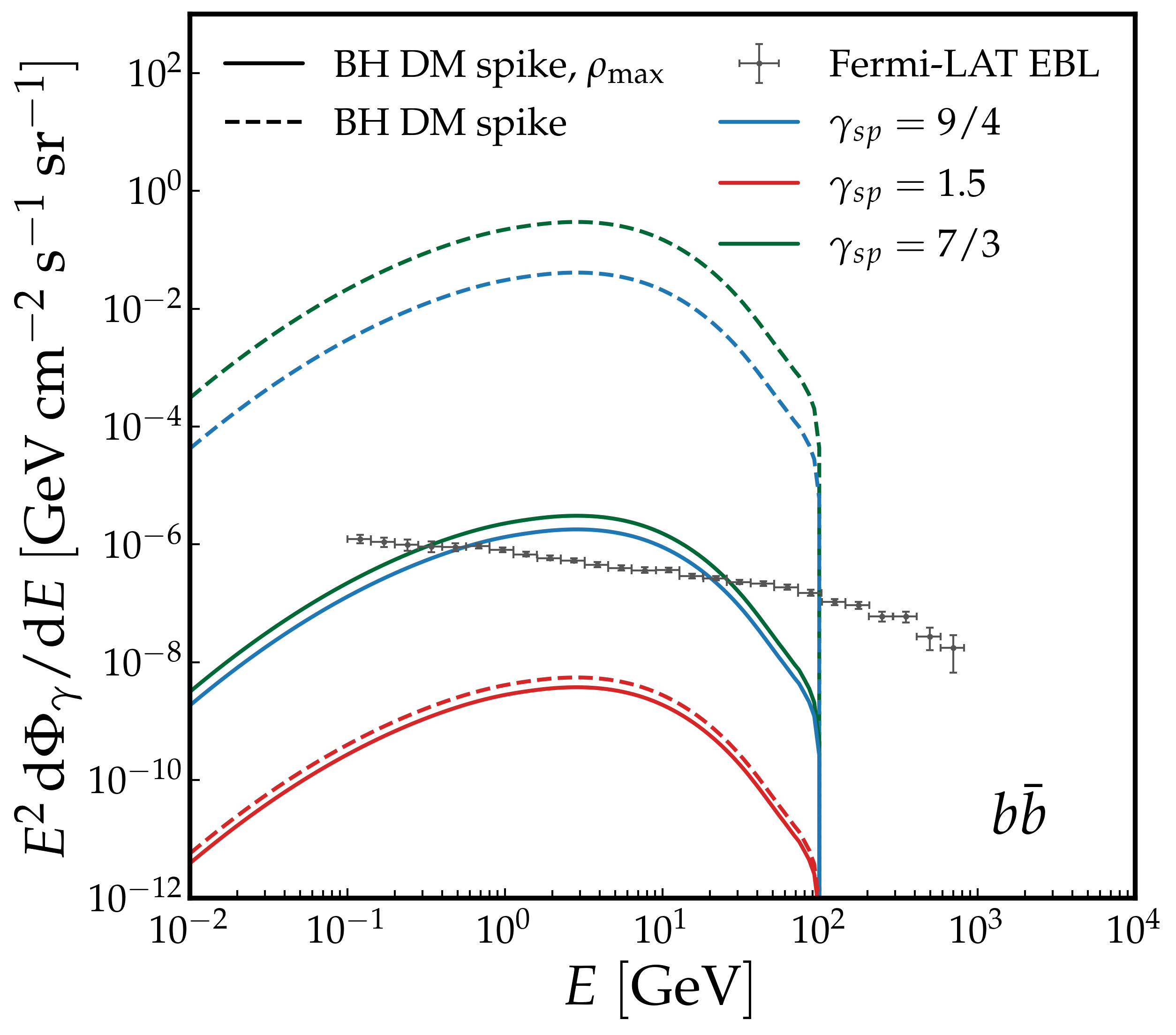}
    \end{subfigure}
    \hfill
    \begin{subfigure}[t]{0.495\textwidth}
        \centering
        \includegraphics[width=\linewidth]{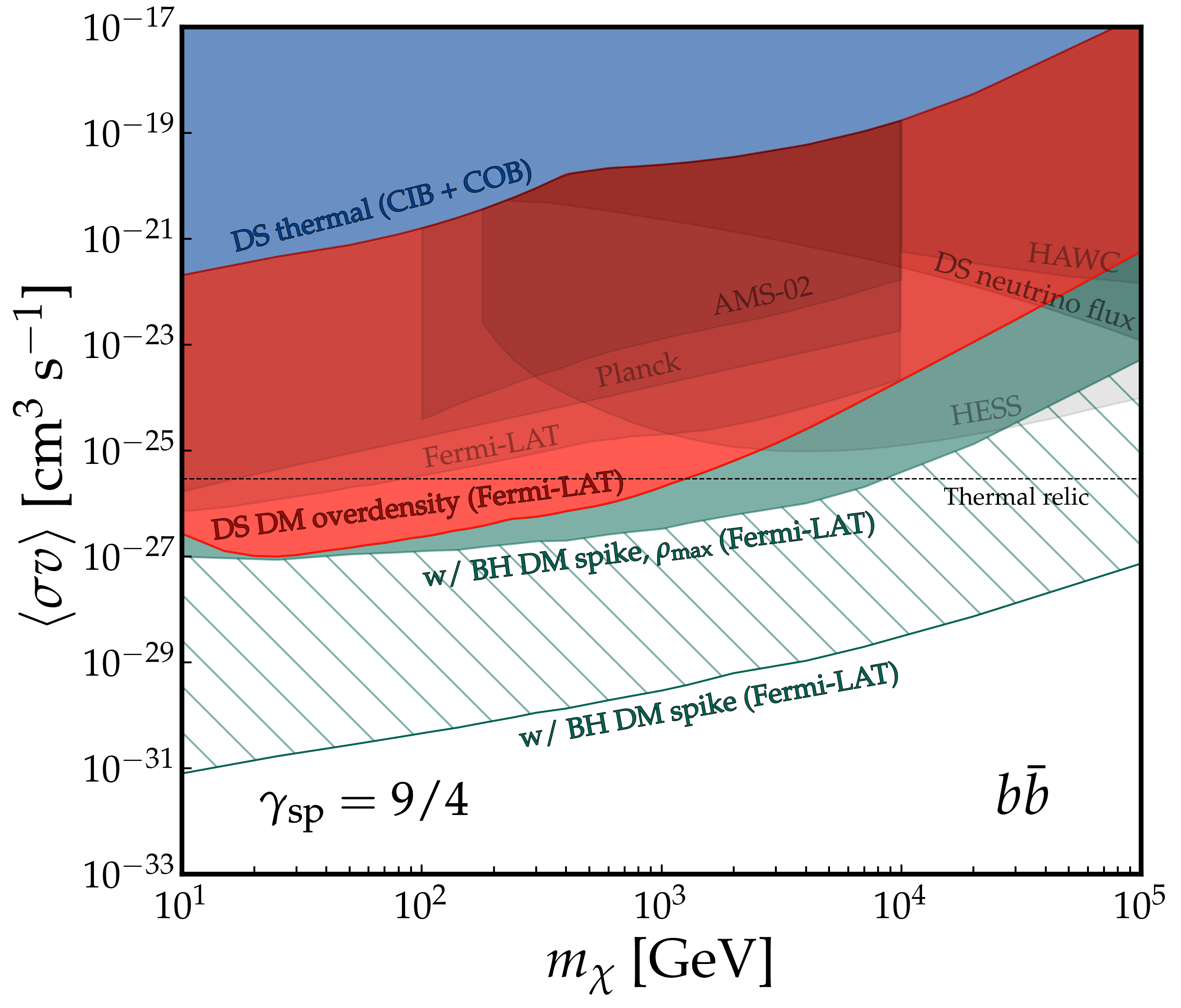}
    \end{subfigure}
\caption{
\textbf{Left:} Diffuse photon flux from BH spikes for different density slopes 
    $\gamma_{\rm sp}=1.5$ (red), $9/4$ (blue), and the fiducial $7/3$ (green).  
    Dashed curves show the results for static spikes while solid curves include density restriction from DM annihilation depletion $\rho_{\max}$.  
    \textbf{Right:} Corresponding constraints on the annihilation cross-section for the 
    $b\bar b$ channel, assuming $\gamma_{\rm sp}=9/4$. The limits are slightly weaker than for 
    the reference $\gamma_{\rm sp}=7/3$ case shown in the main text.
}
    \label{fig:gamma_comp}
\end{figure}

As shown in the left panel of Fig.~\ref{fig:gamma_comp}, a more cored spike with $\gamma_{\rm sp}=1.5$ is only mildly affected by the $\rho_{\max}$ cutoff, since most of the emission originates at radii where the density is well below $\rho_{\max}$.  
In contrast, for the steeper $\gamma_{\rm sp}=9/4$ profile the luminosity is sharply reduced when the cutoff is imposed, reflecting the fact that a substantial fraction of the annihilation rate originates from the innermost region where $\rho>\rho_{\max}$.  
This behavior is broadly similar to the reference $\gamma_{\rm sp}=7/3$ case.

Beyond the choice of $\gamma_{\rm sp}$, BH spikes may evolve significantly over cosmic time.  
As discussed in Sec.~\ref{sec:BH_profile}, processes such as stellar heating, mergers, triaxiality and centrophilic orbits, and departures from perfect adiabatic growth can all alter or erode the spike.  
A more realistic treatment including numerical simulations is well beyond the scope of this work.  
Our comparison of profiles with and without the $\rho_{\max}$ cutoff for several values of $\gamma_{\rm sp}$ therefore serves as a practical benchmark on the plausible ranges of BH-spike luminosities.

The right panel of Fig.~\ref{fig:gamma_comp} shows the resulting constraints for $\gamma_{\rm sp}=9/4$, often considered for secondary infall.  
The bounds become slightly weaker than those obtained for the reference $\gamma_{\rm sp}=7/3$, consistent with the reduced central enhancement of the shallower profile once the cutoff is included.  
For $\gamma_{\rm sp}=1.5$ the BH-spike contribution never exceeds the DS external DM overdensity limits shown in Fig.~\ref{fig:constraints}, hence we do not display additional constraint curves.

\label{app:effects}

\bibliography{refs}
\addcontentsline{toc}{section}{Bibliography}
\bibliographystyle{JHEP}

\end{document}